\documentclass[preprint,12pt,authoryear]{elsarticle}

\usepackage{url}
\usepackage{lineno}
\usepackage[inline]{trackchanges} 
\usepackage{soul}
\usepackage{array}
\usepackage{tcolorbox}
\usepackage{booktabs} 
\usepackage{graphicx}
\usepackage{natbib}
\usepackage{subcaption}
\usepackage{stmaryrd}
\usepackage{bm}
\usepackage{amssymb, amsmath, amsthm}

\journal{Weather and Climate Extremes}

\begin{document}

\begin{frontmatter}



\title{Investigating the Robustness of Extreme Precipitation Super-Resolution Across Climates} 

\author[1]{Louise Largeau}
\author[1,3]{Tom Beucler}
\author[2]{David Leutwyler}
\author[1,3]{Gregoire Mariethoz}
\author[1,4]{Valerie Chavez-Demoulin}
\author[1]{Erwan Koch}

\affiliation[1]{organization={Expertise Center for Climate Extremes, University of Lausanne},
               city={Lausanne},
               postcode={1015},
               state={VD},
               country={Switzerland}}

\affiliation[2]{organization={Federal Office of Meteorology and Climatology MeteoSwiss},
               city={Zurich},
               postcode={8058},
               state={ZH},
               country={Switzerland}}

\affiliation[3]{organization={Faculty of Geosciences and Environment, University of Lausanne},
               city={Lausanne},
               postcode={1015},
               state={VD},
               country={Switzerland}}

\affiliation[4]{organization={Faculty of Business and Economics (HEC), University of Lausanne},
               city={Lausanne},
               postcode={1015},
               state={VD},
               country={Switzerland}}

\begin{abstract}
The coarse spatial resolution of gridded climate models, such as general circulation models, limits their direct use in projecting socially relevant variables like extreme precipitation. Most downscaling methods estimate the conditional distributions of extremes by generating large ensembles, complicating the assessment of robustness under distributional transformations, such as those induced by climate change. To better understand and potentially improve robustness, we propose super-resolving the parameters of the target variable's probability distribution directly using analytically tractable mappings. Within a perfect-model framework over Switzerland, we demonstrate that vector generalized linear and additive models can super-resolve the generalized extreme value distribution of summer hourly precipitation extremes from coarse precipitation fields and topography. We introduce the notion of a ``robustness gap'', defined as the difference in predictive error between present-trained and future-trained models, and use it to diagnose how model structure affects the generalization of each quantile to a pseudo-global warming scenario. By evaluating multiple model configurations, we also identify an upper limit on the super-resolution factor based on the spatial auto- and cross-correlation of precipitation and elevation, beyond which coarse precipitation loses predictive value. Our framework is broadly applicable to variables governed by parametric distributions and offers a model-agnostic diagnostic for understanding when and why empirical downscaling generalizes to climate change and extremes.
\end{abstract}

\begin{graphicalabstract}
\includegraphics[width=\textwidth]{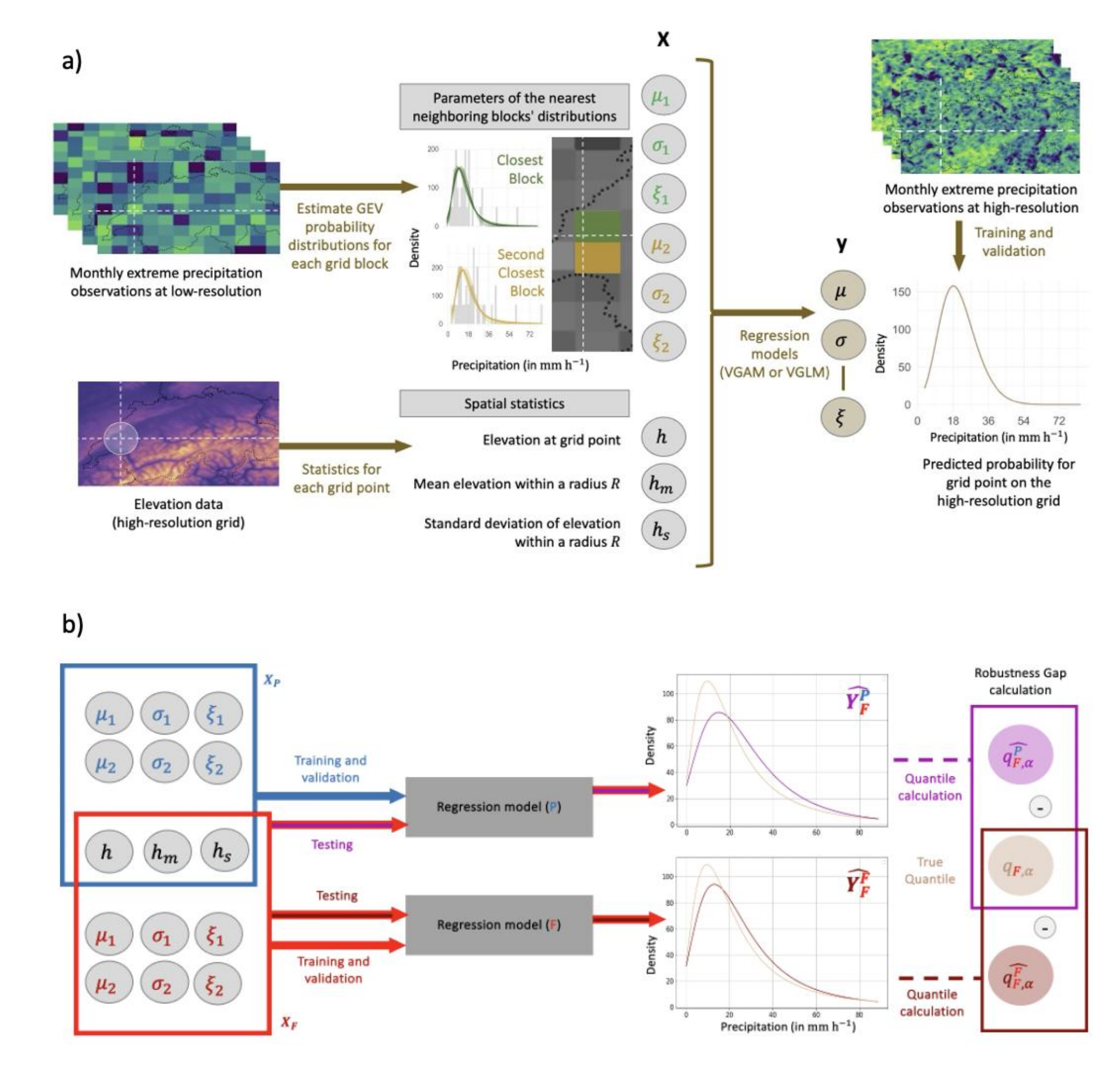}
\end{graphicalabstract}

\begin{highlights}

\item Introduces the concept of super-resolving distributions of weather/climate extremes

\item Super-resolves the GEV distribution of hourly precipitation extremes over Switzerland

\item Uses VGAMs to visualize how GEV parameters vary with covariates via splines

\item Introduces interpretable ``robustness gap'' to explain generalization to climate change

\item Identifies an upper bound on super-resolution factors using spatial statistics

\end{highlights}


\end{frontmatter}


\section{Introduction} 


Climate adaptation requires projecting high-impact weather events at local scales, notably extreme precipitation due to its impacts on ecosystems and infrastructure \citep{fowler2007,gimeno22}. General circulation models (GCMs) are too coarse (with horizontal grid spacing around 100\,km) to explicitly simulate such extremes \citep{maraun2016bias,benestad04}. This limitation warrants the use of expensive regional climate models that can only be run selectively, especially at convection-permitting scales \citep{schar20}, potentially under-sampling projection uncertainty \citep{hawkins2011potential}. Empirical downscaling, including statistical \citep{maraun2018statistical} and machine learning (ML; \citet{rampal2024enhancing}) methods, offers a promising complement or alternative by directly predicting relevant variables at local scales when suitable training data are available. Recent advances in ML-based super-resolution \citep{wang2022comprehensive} and generative modeling \citep{yang2023diffusion} have further fueled the rapid development of empirical downscaling for precipitation \citep{rampal2024reliableGAN,srivastava2024precipitation,Rampal2022interpretabledeeplearning,vandal2019intercomparison}. However, their application to climate change remains limited by challenges in understanding their \textit{robustness}---i.e., how well they extrapolate to warmer climates \citep{hernanz22suitability}. This is especially true for extremes, where stationarity is difficult to assess from historical performance alone \citep{dixon2016evaluating}.

This motivates pseudo-reality experiments (also called ``model as truth'' or ``perfect model''), in which outputs from a dynamical regional climate model are treated as pseudo-observations for empirical downscaling, enabling direct benchmarking of generalization capabilities \citep{maraun2015value}. While such experiments have helped identify best practices (e.g., optimal predictor sets) for improving the robustness of simple statistical downscaling algorithms \citep{charles1999validation,vrac2007general,dayon2015transferability}, they remain under-used for more sophisticated methods. Notable exceptions include \citet{legasa23}, who showed that a posteriori random forests targeting parameters of the precipitation gamma distribution \citep{legasa22aposterioriRF} generalize better to a warmer climate than generalized linear models and convolutional neural networks; \citet{bano24transferability}, who found that deep learning emulators generalize more reliably when GCM predictors are bias-adjusted to the upscaled regional model; and \citet{rampal2024extrapolation}, who showed that generative adversarial networks outperform deterministic convolutional neural networks in projecting warming-driven precipitation extremes. An emerging challenge is that these results remain scattered---typically evaluated per case or quantile---and we still lack a simple framework to explain what drives robustness across varying degrees of extremeness.

To address this gap, we introduce the concept of super-resolution of extremal distributions, a novel approach that focuses on learning to increase the spatial resolution of the parameters governing precipitation extremes rather than reconstructing precipitation fields. This shift from traditional field-based super-resolution is motivated by tractability and theoretical consistency. Fine-resolution and coarse-resolution extremes do not occur simultaneously, but they typically fall within the same temporal block, making them only weakly paired. Such imperfect correspondence makes field-based super-resolution challenging, whereas distributional modeling naturally accommodates it by focusing on aggregated statistical properties rather than individual values. Moreover, super-resolution of distributions is generally not treated on a per-GCM basis. This contrasts with the full empirical downscaling pipeline required for local climate projections, where global and sometimes regional climate model outputs must be bias-corrected \citep{francois2020multivariate,cannon2018multivariate,vrac2015multivariateBC} before being brought to the desired spatial scale. Beyond tractability, working with distributions offers several advantages. It aligns with practical needs for risk assessment at the local level: decision-makers often rely on quantiles or return levels derived from distributions, not raw precipitation fields. It may reduce the risk of overfitting to local patterns and enhances generalization to other regions and climates, as the model learns a small set of interpretable parameters rather than high-dimensional fields. Finally, this approach is computationally efficient, requiring fewer data and enabling interpretable models with lower training costs. \citet{friederichs2010statistical} previously downscaled precipitation block maxima by targeting GEV parameters from large-scale circulation features. However, we focus here on cases where convection-permitting extremes (1--5\,km) are at least partially resolved at coarser resolutions (12--50\,km), consistent with the HighResMIP2 range targeted by next-generation of Earth system models \citep{roberts2025highresmip2}, allowing a super-resolution setup.


Our generative super-resolution model combines regression-based modeling with extreme value theory \citep{vrac2007stochastic}. It learns the parameters of the Generalized Extreme-Value (GEV) distribution governing fine-resolution precipitation maxima from those governing their coarse-resolution counterparts. To ensure interpretability, we use Vector Generalized Additive Models (VGAM) and Vector Generalized Linear Models (VGLM), which allow us to specify a distribution family and examine how features influence predicted distributions through spline terms. This approach offers three key advantages: low training costs, reliable probability estimates, and transparent model behavior.

To quantify our models' robustness to climate change, we adopt a pseudo-reality framework over Switzerland, where complex topography---ranging from alpine areas above 4,000\,m to low-lying regions below 200\,m---drives strong spatial variability in precipitation. Observations show a positive trend in annual maximum daily precipitation at 91\% of Swiss stations, with 10-minute summer precipitation intensities increasing at 5.7\% per decade \citep{bauer2024observedevolutionprecip,scherrer2016trends}. Convection-permitting model ensembles reproduce these changes more faithfully than coarser models and project a 6-7\% increase in heavy summer precipitation intensity per degree of warming, despite an overall summer mean decrease \citep{estermann2025projections,ban2021first}. As summer precipitation is dominated by convective events, it remains particularly challenging to downscale without underestimating variability \citep{zubler2014localized,schmidli2007downscaling}, though this can be partly mitigated by explicitly incorporating temperature dependence in precipitation scaling \citep{moraga2024subdailyrainfall}.

Motivated by the challenge of super-resolving extreme summer precipitation, we base our pseudo-reality experiment on the present-future convection-permitting simulation pair of \citet{Hentgen2019data} (Section~\ref{sec_data}). We then formalize the task of super-resolving distributions and define the ``robustness gap'' to quantitatively assess model generalization across climates (Section~\ref{Sec_Theory}). Applied to the super-resolution of the GEV using VGAMs and VGLMs (Section~\ref{sec_methodology}), our diagnostic framework explains generalization errors across quantiles, interprets model behavior via splines, and identifies performance limits when the super-resolution factor becomes too large (Section~\ref{sec_results}). We conclude in Section~\ref{sec_conclusion}. The Supplementary Material (SM) provides technical derivations that support Sections~\ref{sec_methodology} and~\ref{sec_results} of the manuscript.

\section{Data} 
\label{sec_data}

We super-resolve hourly precipitation model output over Switzerland for 11 European summer (JJA) seasons in both historical (1999–2009) and projected (2079–2089) climates. We focus on summer months to isolate convective precipitation; Swiss summers are characterized by weak synoptic forcing, frequent afternoon thunderstorms, and convective rainfall that are particularly challenging to downscale over the Alps. All data are derived from simulations using the regional weather and climate model COSMO (Consortium for Small-scale Modeling), run over a European domain of approximately $3000 \times 3000$\,km at~2.2\,km grid spacing \citep{Hentgen2019data}. An overview of the simulations is given in Section~\ref{Sec_Simulations}; \citet{Leutwyler2017} documents prior validation of the control run but the associated model data and observations are not part of this study's pipeline. Section~\ref{Subsec_CoarseGrain} describes the construction of the coarse-resolution field from the fine-resolution data. The elevation statistics employed are detailed in Section~\ref{Subsec_ElevSpatialStats}, and the training, validation, and test set configurations are outlined in Section~\ref{Subsec_TrainingValTest}.

\subsection{Simulations}
\label{Sec_Simulations}
We calibrate and test our method using two regional climate simulations at 2.2\,km horizontal resolution from \cite{Hentgen2019data}. Kilometer-scale climate modeling has become increasingly common in recent years \citep{ban2021first, Stevens_2019}, offering several advantages over coarser-resolution simulations. The finer representation of topography and land surface enables more realistic precipitation patterns in complex terrain such as the Alps. In addition, vertical air motion is explicitly resolved by the governing equations, bringing the model formulation closer to physical first principles. Compared to coarser resolutions that require convective parameterization, explicit convection improves the realism of the hydrological cycle, particularly for extreme precipitation and its associated mechanisms and feedbacks \citep{Prein_2020, Schar_2020, Lenderink_2025}. 

To assess generalization, we use model output from a simulation based on the pseudo-global warming (PGW) method \citep{Rasmussen_2011,Adachi_2020,Schar_1996}. PGW aims to simulate a warmer climate by preserving the spatiotemporal structure of historical weather patterns. A regional simulation is first performed using reanalysis-based boundary conditions, referred to as the control simulation (CTRL). CTRL (1999–2008) was observationally validated in \citet{Leutwyler2017} against the ENSEMBLES E-OBS gridded European daily temperature and precipitation dataset \citep{haylock2008EOBS}, the EURO4M Alpine Precipitation Grid Dataset (APGD, \citet{isotta2014APGD}), the Swiss radar-disaggregated hourly precipitation dataset RdisaggH \citep{wuest2010RdisaggH}, the German radar-disaggregated hourly precipitation dataset DisaggDE \citep{paulat2008}, solar surface radiation from the Baseline Surface Radiation Network (BSRN, \citet{hakuba2014}), and lightning from the UK Met Office Arrival Time Difference network (ATDnet, \citet{anderson2014lightning}). 

A physically consistent climate change signal (``climate delta'') is then applied to the CTRL simulation's boundaries, and the simulation is repeated. In \cite{Hentgen2019data}, the deltas are 30-year climatological differences: 2070–2099 (RCP8.5) minus 1971–2000, taken from CMIP5 runs of the Max Planck Institute Earth System Model Low Resolution (MPI-ESM-LR) GCM \citep{kroner2017separating}. These deltas vary with latitude, longitude, elevation, and month. The PGW simulation hence retains the sequence of weather events from CTRL, adjusted by the imposed climate signal. As such, the PGW simulation primarily captures thermodynamic changes (e.g., increases in temperature and moisture) while preserving large-scale circulation patterns from the historical record \citep{Hall_2024}. This reduces confounding effects from circulation biases, providing a dataset of intermediate complexity with a clear RCP8.5 warming signal to stress-test the generalization of statistical downscaling methods.

%
%

\subsection{Coarse-Graining}
\label{Subsec_CoarseGrain}

To ensure comparability across climates, we adopt an idealized super-resolution framework in which the coarse-resolution inputs are coarsened versions of the fine-resolution targets. Hourly precipitation data are aggregated to coarser grid spacings via mean pooling, i.e., using a square spatial filter of prescribed length on the native COSMO grid. To assess robustness across not only climates but also super-resolution factors, we apply pooling lengths of 13.2\,km ($\times$6), 26.4\,km ($\times$12), and 52.8\,km ($\times$24). We then compute monthly maxima on the coarsened low-resolution data (features) and on the fine-resolution data (targets).
\begin{figure}
    \centering
    \includegraphics[width=1\textwidth]{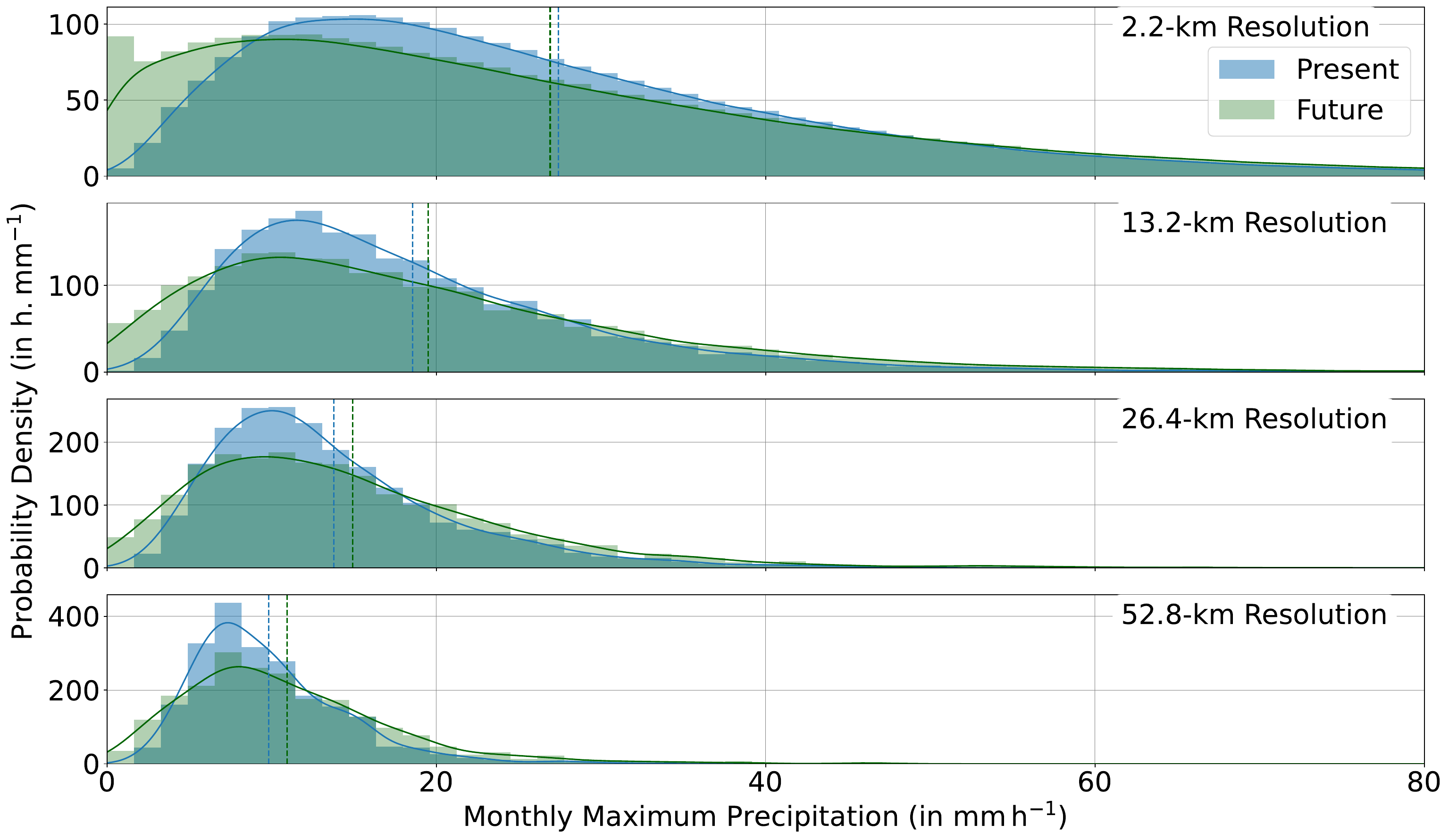} 
    \caption{Contraction of empirical distributions of precipitation extremes in Switzerland with decreasing resolution. Histograms (bars), kernel density estimates (curves), and mean values (dashed vertical lines) for present (blue) and +4K (green) climates at horizontal resolutions of 2.2\,km, 13.2\,km, 26.4\,km, and 52.8\,km.}
    \label{fig_blurring} 
\end{figure}
As shown in Figure~\ref{fig_blurring}, decreasing spatial resolution leads to substantial information loss. Precipitation values become more uniform, with reduced means and variances of extremes consistent with the geostatistical concept of ``change of support'' \citep{onibon2004, no-wook2023}. The highest extremes are particularly affected, with distribution tails shortening as resolution decreases. This is due to the localized nature of extremes: coarse-resolution cells blend multiple events within a grid block, smoothing out their intensity and variability. These patterns remain consistent across all resolutions in both climates. Future climates exhibit broader, flatter distributions with wider supports, heavier tails and higher standard deviations, indicating increased dispersion. 

\subsection{Elevation Spatial Statistics}
\label{Subsec_ElevSpatialStats}

To incorporate elevation as a covariate, we calculate spatial statistics from the digital elevation model used by COSMO. Specifically, we compute two statistics from the 2.2\,km-resolution elevation field $h$ by defining a circular neighborhood of radius $R$ centered on each grid point: the mean elevation $h_{\text{m}}$ and the standard deviation $h_{\text{s}}$, both calculated over the values of $h$ within the circle. The quantities $h$, $h_{\text{m}}$, and $h_{\text{s}}$ are all considered during feature selection, and the radius $R $ is treated as a model hyperparameter.

\subsection{Training, Validation, and Test Split}
\label{Subsec_TrainingValTest}

\begin{figure}
    \centering
    \includegraphics[width=1\textwidth]{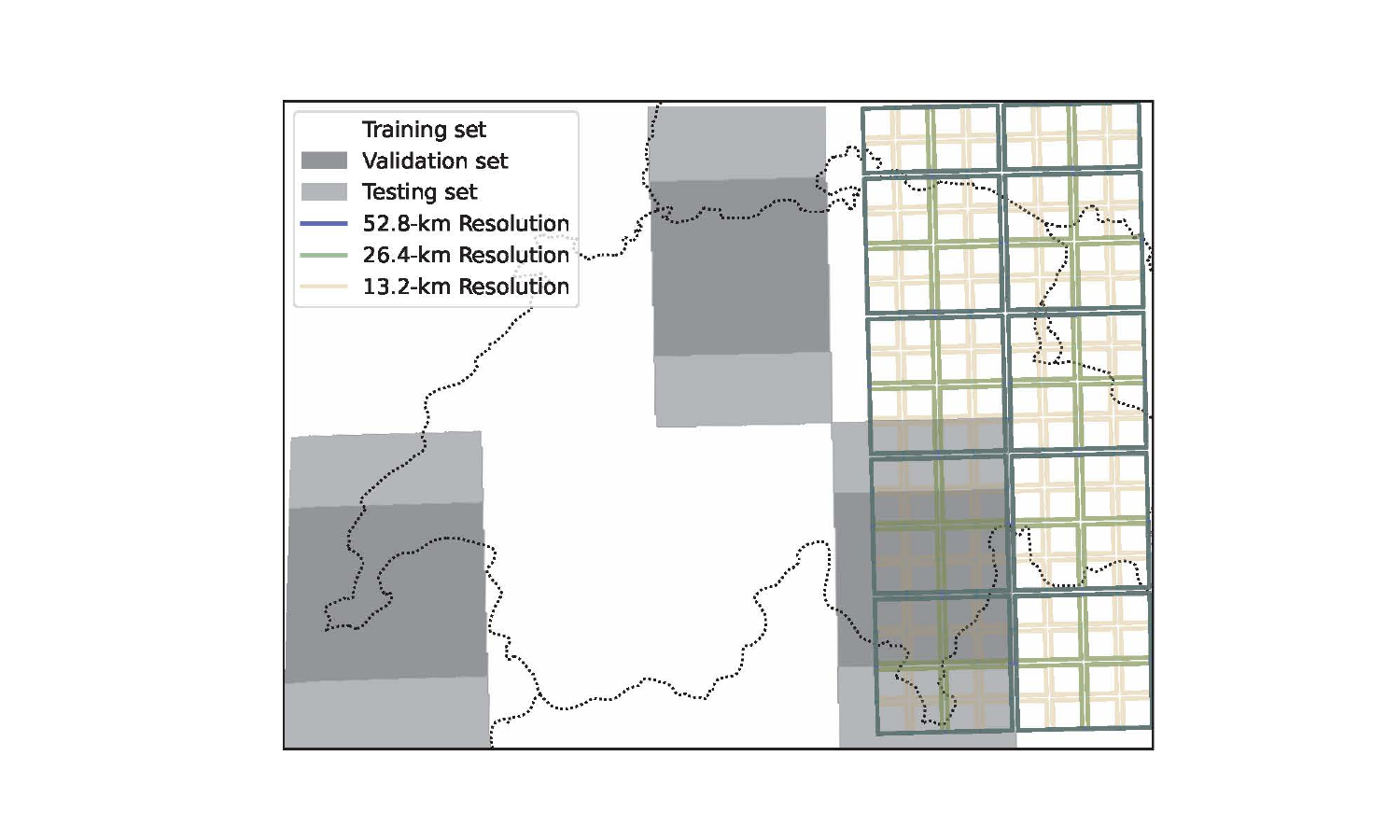} 
    \caption{Spatial partitioning of the domain into training (white), validation (dark gray), and test (light gray) regions. The spatial blocks used to define the coarse-resolution data—13.2\,km (yellow), 26.4\,km (green), and 52.8\,km (blue)—follow the rotated grid of the COSMO climate model.}
    \label{fig_split} 
\end{figure}

To prevent overfitting and to ensure objective model evaluation, we define spatially separated splits \citep{valavi2019,brenning2012}, shown in Figure~\ref{fig_split}. Rather than assigning entire contiguous regions to each split, we partition the domain into 10 spatial regions, which are then distributed across the training, validation, and test sets. This approach preserves spatial separation while maintaining representativeness of the overall data distribution in each set. The training set (white) includes 70\% of the data and is used to optimize the model’s trainable parameters, providing enough information to learn the super-resolution mapping. The validation set (dark gray), comprising 17\% of the data, is used for model and feature selection and helps prevent overfitting. The remaining 13\% forms the test set (light gray), which evaluates the model’s ability to generalize to unseen regions. We refer to the generalization quantified by the test set as ``spatial generalizability'' to distinguish it from the models' ability to generalize across climates.

\section{Theory}
\label{Sec_Theory}

To better understand the super-resolution of extremes and its sensitivity to spatial resolution and climate change, this section presents novel tools to characterize the distribution of extreme events at high spatial resolution, using information from coarser-resolution distributions and auxiliary variables. We begin by introducing and formalizing the concept of super-resolving parametric distributions, which enables the inference of fine-scale extremes from aggregated inputs. We then examine the spatial scales at which coarse-resolution predictors lose their capacity to inform fine-scale extremes. Finally, we propose strategies for assessing the robustness of data-driven models in the context of non-stationary climate conditions.

\subsection{Super-Resolution of Distributions}
\label{subsec_superresolution_dist}

Although super-resolution of physical fields has been widely studied and applied, the concept of \textit{super-resolution of distributions} is, to the best of our knowledge, novel. Access to fine-resolution distributional information is essential for risk assessment and impact studies at fine spatial scales, as it enables the computation of various distributional summaries—such as return periods and exceedance probabilities—that are critical for decision-making under uncertainty.

Let \( X \) denote a physical field—such as precipitation intensity—defined over a spatial domain \( \mathcal{S} \). In statistical terms, \( X \) is a random field, and the value of the field at any site \( \bm{s} \in \mathcal{S} \) is a random variable \( X(\bm{s}) \), whose cumulative distribution function (CDF) is given by
\begin{equation}
F_{X(\bm{s})}(x) \overset{\mathrm{def}}{=} \mathbb{P}(X(\bm{s}) \leq x), \quad x \in \mathbb{R}.
\end{equation}
For simplicity, we assume that this distribution is parametric and characterized by a parameter vector \( \bm{\theta}_{\bm{s}} \in \mathbb{R}^q \).

Consider two spatial grids, \( \mathcal{G}_1 \) and \( \mathcal{G}_2 \), such that the resolution of \( \mathcal{G}_1 \) is coarser than that of \( \mathcal{G}_2 \). We assume that the marginal distribution of \( X \) is known at each grid point \( \bm{s}_1 \in \mathcal{G}_1 \), but unknown at the finer-resolution grid points \( \bm{s}_2 \in \mathcal{G}_2 \). The goal of \textit{super-resolution of distributions} is to infer the distribution of \( X(\bm{s}) \) at each fine-resolution grid point \( \bm{s} \in \mathcal{G}_2 \), using the known distributions at coarse-resolution grid points in \( \mathcal{G}_1 \), possibly along with auxiliary covariates such as topography.

A concrete example for \( \mathcal{G}_1 \) and \( \mathcal{G}_2 \) consists of regular two-dimensional grids defined over the study domain $\mathcal{S}$, with mesh sizes \( \delta_1 > 0 \) and \( \delta_2 > 0 \). 

\begin{equation}
\mathcal{G}^{\delta_1} \overset{\mathrm{def}}{=} \left\{ (\delta_1 \cdot i_1, \delta_1 \cdot i_2) \in \mathcal{S} : (i_1, i_2) \in \mathbb{N}^2 \right\},
\end{equation}
\begin{equation}
\mathcal{G}^{\delta_2} \overset{\mathrm{def}}{=} \left\{ (\delta_2 \cdot i_1, \delta_2 \cdot i_2) \in \mathcal{S} : (i_1, i_2) \in \mathbb{N}^2 \right\},
\end{equation}
where \( \delta_1 > \delta_2 \), so that \( \mathcal{G}^{\delta_2} \) has higher spatial resolution than \( \mathcal{G}^{\delta_1} \).

Let \( \bm{\psi}_1 \) denote the concatenation of all parameter vectors \( \bm{\theta}_{\bm{s}} \) for \( \bm{s} \in \mathcal{G}_1 \). The task of super-resolving distributions then consists in estimating, for each \( \bm{s} \in \mathcal{G}_2 \),
\begin{equation}
\label{Eq_Fundamental_Superres_Dist}
    \bm{\theta}_{\bm{s}} = f(\bm{\psi}_1, \bm{\tau}_{\bm{s}}),
\end{equation}
where \( \bm{\tau}_{\bm{s}} \) represents a set of features specific to grid point \( \bm{s} \) (e.g., latitude, longitude, topography, or land-use features), and \( f \) is a function—possibly learned from data—that maps coarse-resolution distributional information and local covariates to fine-resolution distribution parameters.

\subsection{At Which Level of Coarsening Does Super-Resolution Fail?}
\label{Subsec_Resol_BreakSR}



As spatial resolution decreases, fine-resolution information about the variable of interest is progressively lost due to the homogenization of values across larger grid points. This degradation may cause super-resolution techniques to fail. The objective of this section is to introduce a general and easy-to-implement methodology for identifying the resolution threshold at which super-resolution becomes ineffective—specifically, when the distribution of the variable of interest at low resolution provides little to no insight into its distribution at higher resolution.

For clarity of exposition, we assume the existence of two regular grids, \( \mathcal{G}^{\delta_1} \) and \( \mathcal{G}^{\delta_2} \), as introduced in Section~\ref{subsec_superresolution_dist}, with \( \delta_1 > \delta_2 > 0 \), and where \( \delta_1 \) is an integer multiple of \( \delta_2 \). While this assumption simplifies the presentation, our approach naturally extends to more general settings involving arbitrary grids.

A natural approach for addressing the question posed in this section is to examine the spatial correlation function of the field of interest (e.g., precipitation intensity), denoted by \( X \). This function captures how similar the values of the field are at different sites, depending on their spatial separation—providing insight into the scale and structure of spatial patterns, such as whether events are localized or spread out over larger regions. In particular, it allows us to assess whether the value of the field at a distant site still carries meaningful information about the value at a reference point, which is crucial for identifying the resolution below which super-resolution techniques may no longer yield significant benefits. However, in most cases, the field \( X \) is not stationary, and we therefore standardize it so that it has zero mean and unit variance at each grid point, resulting in a field denoted by \( \tilde{X} \). In practice, we compute, for each \( \bm{s} \in  \mathcal{G}^{\delta_2} \), the temporal mean and standard deviation
$$\bar{X}(\bm{s})\overset{\mathrm{def}}{=}\frac{1}{N} \sum_{i=1}^N X_i(\bm{s}) \qquad \text{and} \qquad \hat{\mathrm{std}}_{X}^{\text{(t)}}(\bm{s}) \overset{\mathrm{def}}{=} \sqrt{\frac{1}{N-1} \sum_{i=1}^N \left( X_i(\bm{s}) - \bar{X}(\bm{s}) \right)^2},$$
and define $\tilde{X}$ as
\begin{equation}
\tilde{X}(\bm{s}) \overset{\mathrm{def}}{=} \frac{X(\bm{s}) - \bar{X}(\bm{s})}{\hat{\mathrm{std}}_{X}^{\text{(t)}}(\bm{s})}, \quad \bm{s} \in  \mathcal{G}^{\delta_2},
\end{equation}
where $N$ denotes the number of time points (equal to $33$ in our application) and $X_i$ is the field $X$ observed at time $i$.
This standardization provides an approximate treatment toward second-order stationarity, as the resulting field \( \tilde{X} \) has constant mean and variance across space. However, this is not a full stationarization: correlations may still depend on absolute gridpoint positions rather than solely on relative displacement.

We denote by \( \rho_{\tilde{X}}(\bm{r}) \), for \( \bm{r} \in  \mathcal{G}^{\delta_2} \), the spatial correlation function of the transformed field. Under the assumption of isotropy, which we adopt here, this function depends only on the distance between sites, and is thus written \( \rho_{\tilde{X}}(r) \). We define
\begin{equation}
B_r \overset{\mathrm{def}}{=} \left\{ (\bm{s}_1, \bm{s}_2) \in  \mathcal{G}^{\delta_2} : d(\bm{s}_1, \bm{s}_2) \in \left[ r - \frac{\Delta r}{2},\, r + \frac{\Delta r}{2} \right] \right\}
\end{equation}
as the set of all pairs of sites whose pairwise distances fall within a ball of radius \( \Delta r \) centered at \( r \), and denote by \( M_r \) the number of such pairs.
Then the spatial correlation function of $\tilde{X}$ is defined by 
\begin{equation}
\label{eq_spatial_correl_normalized}
\rho_{\tilde{X}}(r) \overset{\mathrm{def}}{=}  \frac{1}{N} \sum_{i=1}^N \frac{1}{M_r} 
\sum_{\substack{(\bm{s}_1, \bm{s}_2) \in B_r}} 
\left( \tilde{X}_i(\bm{s}_1) - \langle \tilde{X}_i \rangle \right)
\left( \tilde{X}_i(\bm{s}_2) - \langle \tilde{X}_i \rangle \right)
\Big/ {\mathrm{std}_{\tilde{X}_i}^{\text{(s)}}}^2,
\end{equation}
where
\begin{equation}
\langle \tilde{X}_i \rangle \overset{\mathrm{def}}{=} \frac{1}{M} \sum_{j=1}^M \tilde{X}_i(\bm{s}_j), \quad
{\mathrm{std}_{\tilde{X}_i}^{\text{(s)}}}^2 \overset{\mathrm{def}}{=} \frac{1}{M} \sum_{j=1}^M \left( \tilde{X}_i(\bm{s}_j) - \langle \tilde{X}_i \rangle \right)^2.
\end{equation}
are the spatial mean and the spatial variance of the field $\tilde{X}_i$ observed at time $i$, with $M$ denoting the number of fine-resolution grid points.
In \eqref{eq_spatial_correl_normalized}, the term within the sum over $i$ corresponds to the empirical estimator of the spatial correlation function at distance $r$ for the $i$-th temporal replicate of $\tilde{X}$. These terms are then averaged over time.

Let \( \mathcal{B}^{\delta_1} \) denote a generic block of side length \( \delta_1 \), and let \( \mathcal{B}_{i}^{\delta_2} \) represent the \( i \)-th sub-block of side length \( \delta_2 \) contained within \( \mathcal{B}^{\delta_1} \). One way to define a resolution threshold below which super-resolution becomes ineffective is to identify the smallest \( \delta_1 \), denoted \( \delta_l \), such that the spatial correlation between the most distant sub-blocks within \( \mathcal{B}^{\delta_1} \) falls below a predefined threshold \( \epsilon \) (e.g., 0.1), which may depend on the variable of interest. Formally, we define
\begin{equation}
\label{Eq_resolution_break_SR}
\delta_l \overset{\mathrm{def}}{=} \min \left\{ \delta_1 : \rho_{\tilde{X}} \left( \max_{\mathcal{B}_i^{\delta_2}, \mathcal{B}_j^{\delta_2} \subset \mathcal{B}^{\delta_1}} d(\mathcal{B}_i^{\delta_2}, \mathcal{B}_j^{\delta_2}) \right) \leq \epsilon \right\},
\end{equation}
where \( d(\cdot, \cdot) \) denotes the distance between the barycenters of the respective blocks. The term involving the maximum captures the largest separation between any two sub-blocks of size \( \delta_2 \) within a block of size \( \delta_1 \). If this separation is large enough for the correlation between the blocks to drop below \( \epsilon \), it indicates that the distribution at the coarser resolution \( \delta_1 \) no longer provides meaningful information about the finer-scale distribution at resolution \( \delta_2 \). Since the spatial correlation function \( \rho_{\tilde{X}} \) is typically continuous, \( \delta_l \) can often be characterized as the solution to the equation
\[
\rho_{\tilde{X}} \left( \max_{\mathcal{B}_i^{\delta_2}, \mathcal{B}_j^{\delta_2} \subset \mathcal{B}^{\delta_l}} d(\mathcal{B}_i^{\delta_2}, \mathcal{B}_j^{\delta_2}) \right) = \epsilon.
\]

While we focus here on the spatial correlation function as a measure of dependence, it is important to acknowledge that correlation only captures linear relationships and may not fully represent the complex dependencies associated with extremes. However, by construction, the monthly maxima at lower resolution are obtained through (i) mean pooling of the original hourly data at 2.2\,km resolution and (ii) taking the monthly maxima of these coarse-grained observations (see Section~\ref{Subsec_CoarseGrain} for details). Consequently, the monthly maxima at the lower resolution still involve values at the 2.2\,km scale that are not necessarily extreme, which is why we consider the spatial correlation function an acceptable metric in this context. Nevertheless, the proposed methodology is general and could be applied using alternative dependence measures, such as the extremal coefficient or other metrics tailored to extreme events.

\medskip

Up to this point, we have defined a resolution threshold for the breakdown of super-resolution based on the spatial correlation function of the field of interest. A complementary perspective involves identifying the resolution at which auxiliary variables become more informative than the coarse-resolution representation of the target variable itself. In the case of rainfall, for example, relevant auxiliary variables may include topography, land use, or the dot product between wind vectors and topographic slope (to account for the orientation of terrain relative to prevailing circulation). These variables often retain fine-scale spatial structure by capturing terrain-driven features that strongly influence the underlying physical processes. As a result, they may explain a substantial portion of the variability in extremes of the target variable and, in some cases, offer greater predictive power than the coarse-resolution distribution of the variable of interest.

We denote by \( Y \) the field corresponding to the alternative variable (e.g., topography). We use \( \tilde{X} \) and \( \tilde{Y} \) to denote the potentially normalized versions of \( X \) and \( Y \), respectively. Depending on the context, normalization may or may not be applied to these fields. When the alternative variable \( Y \) is time-invariant---as is the case for topography at the considered timescales---we do not normalize it, and thus set \( \tilde{Y} = Y \). Similarly, when \( X \) represents precipitation and \( Y \) is topography, as in our application, we avoid temporal normalization of \( X \) in order to preserve systematic relationships with elevation and therefore set \( \tilde{X} = X \). To assess the spatial relationship between \( \tilde{X} \) and \( \tilde{Y} \), we compute their spatial cross-correlation function, defined, similarly as in \eqref{eq_spatial_correl_normalized}, by
\begin{equation}
\label{eq_spatial_cross_correl_normalized}
\rho_{\tilde{X}\tilde{Y}}(r) \overset{\mathrm{def}}{=} \frac{1}{N} \sum_{i=1}^N \frac{1}{M_r} 
\sum_{\substack{(\bm{s}_1, \bm{s}_2) \in B_r}} 
\left( \tilde{X}_i(\bm{s}_1) - \langle \tilde{X}_i \rangle \right)
\left( \tilde{Y}_i(\bm{s}_2) - \langle \tilde{Y}_i \rangle \right)
\Big/ \left( \mathrm{std}_{\tilde{X}_i}^{\text{(s)}} \, \mathrm{std}_{\tilde{Y}_i}^{\text{(s)}} \right).
\end{equation}
This function quantifies how strongly the values of the two fields are related across space, depending on their separation.

To determine the resolution at which the alternative variable \( Y \) becomes more informative than the coarse-resolution representation of \( X \), we define
\[
\delta_l \overset{\mathrm{def}}{=} \min \left\{ \delta_1 : \rho_{\tilde{X}} \left( \max_{\mathcal{B}_i^{\delta_2}, \mathcal{B}_j^{\delta_2} \subset \mathcal{B}^{\delta_1}} d(\mathcal{B}_i^{\delta_2}, \mathcal{B}_j^{\delta_2}) \right) \leq\rho_{\tilde{X}\tilde{Y}} \left( \max_{\mathcal{B}_i^{\delta_2}, \mathcal{B}_j^{\delta_2} \subset \mathcal{B}^{\delta_1}} d(\mathcal{B}_i^{\delta_2}, \mathcal{B}_j^{\delta_2}) \right) \right\}.
\]
This expression seeks the smallest block size \( \delta_1 \) such that the spatial correlation of \( \tilde{X} \) between distant sub-blocks becomes lower than the cross-correlation between \( \tilde{X} \) and \( \tilde{Y} \) at the same spatial scale. It identifies the resolution below which the alternative variable \( Y \) provides more useful information about the fine-resolution structure of \( X \) than \( X \) itself at low resolution. As before, since $\rho_{\tilde{X}, \tilde{Y}}$ is continuous, $\delta_l$ is typically the solution of 
$$ \rho_{\tilde{X}} \left( \max_{\mathcal{B}_i^{\delta_2}, \mathcal{B}_j^{\delta_2} \subset \mathcal{B}^{\delta_1}} d(\mathcal{B}_i^{\delta_2}, \mathcal{B}_j^{\delta_2}) \right) = \rho_{\tilde{X}\tilde{Y}} \left( \max_{\mathcal{B}_i^{\delta_2}, \mathcal{B}_j^{\delta_2} \subset \mathcal{B}^{\delta_1}} d(\mathcal{B}_i^{\delta_2}, \mathcal{B}_j^{\delta_2}) \right).$$

\subsection{Quantifying Generalizability and Robustness across Climates}
\label{Subsec_GeneralizabilityRobustnessCC}

The overall goal of this section is to (i) formalize the notions of generalization and robustness abilities of a model in a climate change context; (ii) propose a concrete solution tailored to estimate the quantiles of fine-resolution distributions. To do this, we consider three strategies: (i) a model trained on present-day data and applied to present conditions; (ii) a model trained on present-day data and applied to future conditions; and (iii) a model both trained and evaluated on future data.

 Throughout this section, for notational simplicity, variables may refer either to observations at a specific grid point or to the entire field across all grid points, as should be clear from the context. Let $y$ denote an arbitrary characteristic of the distribution under consideration (e.g., the mean or a quantile). Let \( y_{\mathrm{P}}^{\mathrm{P}} \) be the predicted quantity for the present climate using a model trained on present-day data, \( y_{\mathrm{F}}^{\mathrm{P}} \) be the predicted quantity for the future climate using a model trained on present-day data, and \( y_{\mathrm{F}}^{\mathrm{F}} \) be the predicted quantity for the future climate using a model trained on future data. Moreover, let \( y_{\mathrm{P}} \) and \( y_{\mathrm{F}} \) denote the observed (true) quantities in the present and future climates, respectively. Let \( \ell \) be a generic point-wise loss function that measures the discrepancy between predicted and true values. A common example is the squared error, defined by $
\ell(y, \hat{y}) = (y - \hat{y})^2$.

\subsubsection{Robustness Gap and Normalization Strategies}

We can now define the generalization gap (also referred to as extrapolation gap in our context), as
\begin{equation}
\mathrm{GG}\overset{\mathrm{def}}{=}\left \langle\ell \left( y_{\text{P}}, y_{\text{P}}^{\text{P}} \right)\right \rangle - \left \langle  \ell \left(y_{\text{F}}, y_{\text{F}}^{\text{P}} \right) \right \rangle,
\end{equation}
which quantifies the performance gap between a model trained and evaluated on present-day data versus the same model evaluated on future data. We recall that the notation $\langle \cdot \rangle$ denotes the spatial mean, i.e., the average of the field over all grid points. The use of this operator enables the integration of pointwise losses over the entire
grid. It captures both the inadequacy of parameters learned under current climate conditions when applied to future climates, and the impact of shifts in the model's covariates over time. Thus, this generalization gap can be challenging to interpret, as a high value---indicating poor performance under future climate conditions---may primarily result from shifts in covariates, reflecting fundamentally different environmental conditions, rather than a failure of the model itself. 

To overcome this, we introduce the notion of robustness gap, which is defined by
\begin{equation}
\mathrm{RG}\overset{\mathrm{def}}{=} \left \langle\ell \left(y_{\text{F}}, y_{\text{F}}^{\text{P}} \right) \right \rangle -\left \langle\ell \left(y_{\text{F}}, y_{\text{F}}^{\text{F}}\right) \right \rangle.
\label{eq_gen_def_robustnessgap}
\end{equation}
This metric compares two models evaluated on the same future data, effectively removing the influence of covariate shifts. It quantifies how well a model trained on present-day data performs under future conditions, using a model trained on future data as reference, thereby serving as an indicator of its transferability across climates. A low value indicates that the model parameters are robust to climate shifts, whereas a high one suggests that a model trained on present-day data may not be directly applicable under future climate conditions.

However, the robustness gap defined in~\eqref{eq_gen_def_robustnessgap} is expressed in absolute terms, which makes its interpretation dependent on the scale of the loss function and of the variable under consideration. This lack of normalization complicates comparisons across different metrics, variables, or models. For instance, a robustness gap of 0.5 may indicate negligible degradation for one variable but severe deterioration for another, depending on the underlying scale. To address this limitation, we introduce normalized versions of the robustness gap that are more interpretable.

We consider two normalization strategies. Our first version of the normalized robustness gap is defined as
\begin{equation}
\mathrm{NRG}_{\text{ratio}}
=
\frac{\left \langle \ell \left(y_{\text{F}}, y_{\text{F}}^{\text{P}}\right)\right \rangle
      - \left \langle \ell \left(y_{\text{F}}, y_{\text{F}}^{\text{F}}\right)\right \rangle}
     {\left \langle \ell\left(y_{\text{F}}, y_{\text{F}}^{\text{F}}\right)\right \rangle}.
\label{eq:nrg_ratio}
\end{equation}
Here, the denominator corresponds to the error of the model trained on future data, which serves as a theoretical benchmark. This formulation normalizes the average degradation by this benchmark error, providing an intuitive interpretation: a value of zero indicates no degradation, a value of one means the error doubles compared to the future-trained model, and values greater than one correspond to larger deterioration. Its main advantage lies in its simplicity and interpretability. However, this normalization can be sensitive to extreme pointwise benchmark errors because it relies on a single global mean, which is inherently non-robust to outliers.

Our second version of the normalized robustness gap is
\begin{equation}
\mathrm{NRG}_{\text{pointwise}}
= \left \langle
\frac{\ell\left(y_{\text{F}}, y_{\text{F}}^{\text{P}}\right)
      - \ell\left(y_{\text{F}}, y_{\text{F}}^{\text{F}}\right)}
     {\ell\left(y_{\text{F}}, y_{\text{F}}^{\text{F}}\right)}
\right \rangle.
\label{eq:nrg_pointwise}
\end{equation}
In this case, the degradation relative to the future-trained model is computed for each grid point before averaging. This approach captures local variability and reduces the bias introduced by using a single global normalization factor, although it remains sensitive to extreme pointwise benchmark errors, which propagate into the expectation. Its interpretation is less straightforward than in the first case. In summary, the ratio of expectations in~\eqref{eq:nrg_ratio} emphasizes global interpretability while the expectation of pointwise ratios in~\eqref{eq:nrg_pointwise} prioritizes statistical robustness.

All the metrics presented in this section (generalization gap, robustness gap, and normalized versions of the robustness gap) are computed for a specific dataset combining present-day and future conditions. They are inherently dataset-dependent and may differ across alternative datasets. To account for this variability, one could consider their expected values under the joint distribution of the variables appearing in the corresponding expressions (e.g., $y_{\text{F}},\, y_{\text{F}}^{\text{P}},\, y_{\text{F}}^{\text{F}}$ in the case of the robustness gap).

\subsubsection{Special Case: Robustness Gap for Quantile Predictions}

We now focus on the specific case where the previously defined quantity \( y \) corresponds to a quantile at a generic level \( \alpha \in (0, 1) \). Quantiles are particularly relevant in the context of risk assessment, as they are directly linked to the concept of return periods commonly used in environmental sciences, especially hydrology. The \( T \)-year return level \( z_T \), defined as the level expected to be exceeded on average once every \( T \) temporal units, corresponds to the \( \alpha \)-quantile \( q_\alpha \) of the fitted distribution with \( \alpha = 1 - 1/T \). Return levels provide an interpretable and widely used risk metric; for example, the 100-year return level represents the rainfall intensity expected to be exceeded, on average, once per century.

Let \( q_{\text{F}, \alpha}^{\text{P}} \) denote the predicted quantile for the future climate obtained using a model trained on present-day data, and \( q_{\text{F}, \alpha}^{\text{F}} \) the predicted quantile obtained using a model trained on future data. The observed (true) quantiles in the present and future climate are denoted by \( q_{\text{P}, \alpha} \) and \( q_{\text{F}, \alpha} \), respectively.
In this context, the robustness gap introduced above becomes 
\begin{equation}
\label{Eq_RobustnessGapQuantiles}\mathrm{RG}=\left \langle\ell(q_{\text{F}, \alpha}, q_{\text{F}, \alpha}^{\text{P}}) - \ell(q_{\text{F}, \alpha}, q_{\text{F}, \alpha}^{\text{F}})\right \rangle.
\end{equation}

To derive an explicit and interpretable expression for the robustness gap, we henceforth adopt the pinball loss (also known as quantile loss) to evaluate quantile predictions at level $\alpha \in (0, 1)$. Given a predicted quantile $\hat{q}_{\alpha}$ and an observed (true) value $q_{\alpha}$, the loss is defined as
\begin{equation}
    \ell_{\alpha}(q_{\alpha}, \hat{q}_{\alpha}) \overset{\mathrm{def}}{=} 
    \begin{cases}
        \alpha (q_{\alpha} - \hat{q}_{\alpha}), & \text{if } q_{\alpha} \geq \hat{q}_{\alpha} \\
        (1 - \alpha)(\hat{q}_{\alpha} - q_{\alpha}), & \text{if } q_{\alpha} < \hat{q}_{\alpha}
    \end{cases}
    = (\alpha - \mathbb{I}_{\{q_{\alpha} < \hat{q}_{\alpha} \}})(q_{\alpha} - \hat{q}_{\alpha}),
\end{equation}
where $\mathbb{I}_{\{ \}}$ denotes the indicator function that equals $1$ if the condition in the subscript is satisfied and $0$ otherwise.
Although widely used---particularly in machine learning---this loss function may appear less intuitive than alternatives such as the squared loss, which could also have been employed. Nevertheless, this asymmetric loss function is particularly well-suited for evaluating quantile predictions, as it imposes different penalties for over- and under-predictions depending on the quantile level $\alpha$. Specifically, underestimations are penalized more heavily for high quantiles, while overestimations incur greater penalties for low quantiles. In addition, this loss function facilitates a straightforward decomposition of the robustness gap, enabling a more interpretable analysis of model performance.

We now introduce two interpretable quantities that will naturally appear in the decomposition of the pointwise robustness gap (see below).
Let $\varepsilon_{\text{F}, \alpha} = q_{\text{F}, \alpha}^{\text{F}} - q_{\text{F}, \alpha}$, which is the \text{fit bias} of the model trained on the future climate for the quantile at level $\alpha$, and $\Delta_{\alpha} = q_{\text{F}, \alpha}^{\text{P}} - q_{\text{F}, \alpha}^{\text{F}}$, as it quantifies the sensitivity of the predicted quantile to the choice of training data (whether from the present-day or future climate). We expect $\varepsilon_{F, \alpha}$ to be small if the model is adequate and has been properly trained. With these notations, it is straightforward to obtain that 
\begin{equation}
    q_{\text{F}, \alpha}^{\text{F}}  = q_{\text{F}, \alpha} + \varepsilon_{\text{F}, \alpha}, \quad q_{\text{F}, \alpha}^{\text{P}} = q_{\text{F}, \alpha} + \varepsilon_{\text{F}, \alpha} + \Delta_{\alpha},
\end{equation}
and it can be shown that (see SM, Section A) that the pointwise robustness gap can be written
\begin{align}
\label{Eq_1stExpressions_RG}
\mathrm{PRG} &
\overset{\mathrm{def}}{=}
\ell_\alpha \left(q_{\text{F}, \alpha}, q^{\text{P}}_{\text{F}, \alpha} \right) - \ell_\alpha\left(q_{\text{F}, \alpha}, q^{\text{F}}_{\text{F}, \alpha}\right) \notag \\
&= \Delta_{\alpha} \left( \mathbb{I}_{\{\varepsilon_{\text{F}, \alpha} > -\Delta_{\alpha}\}} - \alpha \right) + \varepsilon_{\text{F}, \alpha} \left( \mathbb{I}_{\{\varepsilon_{\text{F}, \alpha} > -\Delta_{\alpha}\}} - \mathbb{I}_{\{\varepsilon_{\text{F}, \alpha} > 0\}} \right) .
\end{align}
Thus, this difference simplifies to the sum of two terms: one involving the product of $\Delta_{\alpha}$ with a number in $(-1, 1)$ and the other involving the product of $\varepsilon_{\text{F}, \alpha}$ with an indicator function difference that takes values in $\{ -1, 0, 1\}$. Provided the model is suited and well calibrated, $\varepsilon_{\text{F}, \alpha}$ is small and so is the second term. 

Two interesting limiting cases emerge. If $\Delta_{\alpha} = 0$ (i.e., $q_{\text{F}, \alpha}^{\text{P}} = q_{\text{F}, \alpha}^{\text{F}}$), then the pointwise robustness gap is zero for any fit bias $\varepsilon_{\text{F}, \alpha}$.
If $\varepsilon_{\text{F}, \alpha} = 0$ (i.e., the future-trained model is perfectly calibrated), then 
    \begin{equation}
    \mathrm{PRG} =
    \begin{cases}
        (1 - \alpha)\Delta_{\alpha}, & \Delta_{\alpha} > 0, \\
        \alpha |\Delta_{\alpha}|, & \Delta_{\alpha} < 0,
    \end{cases}
    \end{equation}
    showing that the degradation in robustness is directly proportional to the quantile shift and exhibits asymmetry with respect to $\alpha$.

In the case where the object of interest is a parametric distribution, it is useful to investigate which parameters are primarily responsible for the model's lack of robustness under a climate shift. This can be achieved by expressing $\Delta_{\alpha}$ as below. Let us assume that we have a parametric distribution with vector $\bm{\theta} = \left( \bm{\theta}_1, \ldots, \bm{\theta}_q\right)^{\prime}$, where $^{\prime}$ denotes transposition. Then, \( \Delta_{\alpha} \) can be decomposed in terms of the individual gaps coming from each parameter: 
\begin{equation}
\Delta_{\alpha} = \sum_{k=1}^{q} \frac{\partial q_{\alpha}}{\partial \bm{\theta}_k} (\bm{\theta}_k^\text{F} - \bm{\theta}_k^\text{P}) + \text{residual},
\end{equation}
where \( q_{\alpha} \) denotes the predicted \( \alpha \)-quantile of the distribution, and ${\partial q_{\alpha}}/{\partial \bm{\theta}_k}$ reflects the sensitivity of the quantile to variations in that parameter. 

When incorporated into \eqref{Eq_1stExpressions_RG}, this decomposition establishes a general framework for analyzing the contribution of individual parameter shifts to the pointwise robustness gap. It maintains the generality of the expression, ensuring applicability across a wide range of parametric distributions.
Overall, our framework provides a comprehensive understanding of the robustness gap—clarifying not only when the super-resolution model generalizes effectively, when its performance begins to degrade, and when it ultimately fails, but also uncovering the underlying factors driving these behaviors.

\section{Methodology}
\label{sec_methodology}




This section introduces the methodological tools necessary to apply the theoretical framework developed in Section~\ref{Sec_Theory} to rainfall extremes. We implement the super-resolution framework described in Section~\ref{subsec_superresolution_dist} within the context of the Generalized Extreme-Value (GEV) distribution, which is well-suited for the statistical modeling of maxima. 

Section~\ref{Subsec_GEV} formally presents the GEV distribution. We then define, in Section~\ref{Subsec_VGAM_VGLM}, the super-resolution function \( f \) appearing in~\eqref{Eq_Fundamental_Superres_Dist}, which maps the characteristics of the fine-resolution distributions from the parameters of the coarse-resolution distributions and a set of auxiliary features. Finally, Section~\ref{Subsec_Feature_Select} outlines the procedures used for feature selection and hyperparameter tuning to ensure optimal performance of the super-resolution model.

\subsection{Generalized Extreme-Value Distribution}
\label{Subsec_GEV}

To characterize extreme hourly precipitation, we adopt the Generalized Extreme-Value framework from Extreme-Value Theory (EVT), a branch of statistics focused on modeling the behavior of distribution tails. Unlike classical methods that describe central tendencies such as means and variances, EVT provides asymptotically justified models for block maxima or threshold exceedances, making it particularly well suited for assessing the risk of rare, high-impact events like extreme hourly rainfall \citep{GEV}.

Let \( M_n = \max\{Z_1, Z_2, \dots, Z_n\} \) denote the sample maximum of a sequence of random variables \( Z_1, Z_2, \dots, Z_n \). The subscript may represent, for example, the time index in hours, with each variable corresponding to the rainfall amount measured during the preceding hour. In this context, \( M_n \) represents the maximum hourly rainfall observed over a period of \( n \) hours.
Under fairly mild conditions, it is known that, for sufficiently large \( n \), the sample maximum \( M_n \) approximately follows the GEV distribution, whose CDF is
\begin{equation}
F_{\mathrm{GEV}}(z) = \left\{
    \begin{array}{ll}
        \exp \left ( - \left [ 1 + \xi \left( \frac{z - \mu}{\sigma} \right) \right ]^{-\frac{1}{\xi}} \right ) & \mbox{for } \xi \neq 0,\\
        \exp ( - \exp \left( \frac{z - \mu}{\sigma} \right) ) & \mbox{for } \xi = 0,
    \end{array}
\right.
\label{GEV_CD}
\end{equation}
and defined on the set \{$z: 1 + \xi (z - \mu)/\sigma > 0$\} with $\mu \in \mathbb{R}$, $\sigma>0$ and $\xi \in \mathbb{R}$. The location parameter $\mu$ shifts the distribution along the real line, the scale parameter $\sigma$ controls the dispersion, and the shape parameter $\xi$ governs the heaviness of the tail. 
High-intensity hourly precipitation events occur in the far right tail of the rainfall distribution, where observational data are sparse and extreme values disproportionately drive impacts.

This key result underpins the block maxima method, which partitions a time series into non-overlapping blocks, extracts the maximum value from each block, and fits a GEV distribution to the resulting maxima—typically using maximum likelihood estimation; see~(S1) in the SM for the detailed expression of the log-likelihood.


In this study, the GEV distribution is independently fitted at each grid point for both fine-resolution and coarsened (coarse-resolution) datasets. The fine-resolution GEV distribution serves as the reference, while the coarse-resolution distributions provide the baseline for evaluating our approach. A key limitation is the relatively short data record, with only 11 years available for both present-day and future climate scenarios. To increase the number of maxima considered in the block-maxima approach, we treat the monthly maxima from June, July, and August as independent realizations of extremes of a typical summer month, thereby increasing the number of extreme values per grid point from 11 to 33. To ensure temporal independence among these monthly maxima, a minimum separation of five days between events is enforced. While we adopt a block-maxima framework, alternative approaches, such as the \textit{r}-largest order statistics, could be explored in future work.




\subsection{Incorporating Features in the GEV Parameters}
\label{Subsec_VGAM_VGLM}


We now specify the form of the function \( f \) introduced in~\eqref{Eq_Fundamental_Superres_Dist}, adapted to our specific setting in which the underlying parametric distribution is the GEV distribution.
To flexibly capture how the features (parameters of the GEV distributions at coarse-resolution and auxiliary features) influence each parameter of the GEV distribution, we use the Vector Generalized Additive Model (VGAM) framework \citep{Yee2015}. The VGAM extends the familiar Generalized Additive Model (GAM) by allowing for multiple response variables. In our case the vector of response variables is composed of the GEV parameters. 

The GEV parameters at the $i$-th grid point, for $i = 1, \ldots, M$, are linked to features $\bm{x}_i = (x_{i1}, \dots, x_{ip})^{\prime}$ by
\begin{equation}
\label{eq:vgamGEV}
\begin{aligned}
\mu_i &= \eta_1(\bm{x}_i)
       = \beta_{\mu} + \sum_{k=1}^p f_{\mu,k}(x_{ik})\,,\\
\log \sigma_i &= \eta_2(\bm{x}_i)
             = \beta_{\sigma} + \sum_{k=1}^p f_{\sigma,k}(x_{ik})\,,\\
\xi_i &= \eta_3(\bm{x}_i)
        = \beta_{\xi} + \sum_{k=1}^p f_{\xi,k}(x_{ik})\,,
\end{aligned}
\end{equation}
where $\log$ denotes the natural logarithm; $\beta_{\mu}$, $\beta_{\sigma}$, and $\beta_{\xi}$ are the intercepts for the parameters $\mu$, $\sigma$, and $\xi$; and $f_{\mu,k}(\cdot)$, $f_{\sigma,k}(\cdot)$, and $f_{\xi,k}(\cdot)$ are potentially smooth functions of the $k$-th feature, typically represented using basis expansions (e.g., splines). The log-link on $\sigma_i$ ensures positivity of the scale parameter. In the case where all $f_{.,k}$ are linear, the class of VGAMs reduces to the subclass of so-called Vector Generalized Linear Models (VGLMs).  

 We typically model $\mu$ and $\log\sigma$ as smooth functions of the features but, for stability and identifiability, take $\xi$ to be constant (i.e.,\ $f_{\mu,k}\equiv0$ for all $k$). Indeed the estimation of $\xi$ is notoriously imprecise even with large samples of block maxima and allowing $\xi$ to vary with covariates often yields unstable fits \citep{GEV,DavisonSmith1990}.  Moreover, extreme events are inherently rare, so there is typically insufficient data within each covariate ``slice" to support a reliable smooth trend in $\xi$, and attempts to do so can lead to over‐parameterization and degraded predictive performance \citep{EastoeTawn2009}.  Finally, empirical comparisons indicate that introducing time‐ or covariate‐dependence in the location (and sometimes scale) parameters captures the bulk of observed non‐stationarity, while varying $\xi$ provides only marginal improvements at the cost of substantially increased uncertainty \citep{Katz2002,GEV}.

We estimate all unknown quantities by maximizing a penalized log-likelihood (see (S2) in the SM), which balances model fit and smoothness. The penalization discourages overfitting by controlling the complexity of the smooth functions, ensuring that the estimated relationships remain interpretable and generalize well to unseen data.

\subsection{Selecting Features and Hyperparameters}
\label{Subsec_Feature_Select}

Covariate selection is performed using a forward selection procedure, guided by model performance on the hold-out validation set described in Section~\ref{Subsec_TrainingValTest}. Since our objective is to accurately model the fine-resolution distribution, model fit is evaluated using a statistical distance between the modeled and empirical distributions at each grid point.

We adopt the Cramér–von Mises (CVM) distance, a robust metric that  quantifies the discrepancy between two CDFs. For a given grid point, let \( F_N \) denote the empirical CDF derived from the observed sample \( x_1, \ldots, x_N \), and let \( F(x, \bm{\theta}) \) represent the theoretical CDF parameterized by \( \bm{\theta} \). The CVM statistic is defined as
\begin{equation}
D_{\mathrm{CVM}}(F_N, F) \overset{\mathrm{def}}{=} \int \left(F_N(x) - F(x, \bm{\theta}) \right)^2 \mathrm{d}F(x).
\end{equation}
It integrates the squared difference between empirical and modeled CDFs across the entire distribution, providing a sensitive and stable assessment of overall distributional fit.

In practice, we employ a computationally efficient approximation:
\begin{equation}
D^\star_{\mathrm{CVM}}(F_N, F) = \frac{1}{12N} + \sum_{i=1}^{N} \left[F(x_{(i)}, \bm{\theta}) - \frac{2i - 1}{2N}\right]^2,
\end{equation}
which assumes continuity of the theoretical distribution \( F(x, \bm{\theta}) \). Here, the term \( (2i - 1)/(2N) \) corresponds to the expected CDF values under a uniform distribution, serving as a reference for comparison. The squared differences quantify the deviation of the theoretical CDF from this uniform benchmark at the observed data points.

Since we work within the GEV family, this formula can be adapted by explicitly using the known analytical form of the CDF; see~\eqref{GEV_CD}. The total CVM score is computed by summing the individual CVM distances across all grid points in the validation set and, finally, covariates yielding the lowest aggregate CVM score are selected.

A preliminary step is conducted to assess the most informative topographic covariates. For this purpose, elevation-based features are calculated from the digital elevation model using circular neighborhoods of varying radii
$R \in \{ 10, 20, \dots, 100\}$\,km. For each radius, spatial statistics such as the mean and standard deviation of elevation are extracted, and forward selection is applied exclusively to this subset. The results indicate that covariates derived using a 50\,km radius yield the best model fit as measured by the CVM criterion. Therefore, these covariates are retained for inclusion in the full feature selection procedure. 

Finally, we did not include temporal trend terms in the model because they were not statistically significant (see Section D in the SM for details).

\section{Results} 
\label{sec_results}

We first evaluate models trained in the present (CTRL) climate that super-resolve $13.2$\,km inputs to $2.2$\,km targets. We then assess how performance changes as inputs are coarsened to $26.4$ and $52.8$\,km (super-resolution factors of $12\times$ and $24\times$) instead of $13.2$\,km, and explain the observed behavior using the tools introduced in Section~\ref{Subsec_Resol_BreakSR}. Finally, we investigate, using the methodology developed in Section~\ref{Subsec_GeneralizabilityRobustnessCC}, the robustness of our super-resolution models in a warmer climate. 

\subsection{Performance in the Reference Climate}

Our best-performing model for super-resolving precipitation distributions from 13.2\,km to 2.2\,km is a VGAM using eight features. These include two topographic features---elevation ($h$) and its local average ($h_{\text{m}}$)---and six features derived from the coarse-resolution GEV distributions: the location ($\mu_1$), scale ($\sigma_1$), and shape ($\xi_1$) parameters of the nearest block, along with the location ($\mu_2$) and scale ($\sigma_2$) of the second-nearest block.

The VGAM yields interpretable expressions (see \eqref{eq:vgamGEV}) for the parameters of the fine-resolution GEV distribution:
\begin{equation}
\begin{cases}
    \mu(\bm{x}) = \beta_{\mu} + f_{\mu, h}(x_h) + f_{\mu, h_{\text{m}}}(x_{h_{\text{m}}}) + f_{\mu, \mu_1}(x_{\mu_1}) + f_{\mu, \mu_2}(x_{\mu_2}) + f_{\mu, \xi_1}(x_{\xi_1}), \\
    \log(\sigma(\bm{x})) = \beta_{\sigma} + f_{\sigma, \sigma_1}(x_{\sigma_1}) + f_{\sigma, \sigma_2}(x_{\sigma_2}) + f_{\sigma, \mu_1}(x_{\mu_1}), \\
    \xi(\bm{x}) = \beta_{\xi},
\end{cases}
\label{equa_VGAMmodel}
\end{equation}
where $\bm{x}$ represents the covariates at a specific grid point. Therefore, at each grid point, the parameters of the target GEV distribution ($\hat{\mu}$, $\hat{\sigma}$, and $\hat{\xi}$) are obtained by evaluating each spline at the corresponding covariate value at that point, summing all contributions, and adding an intercept.  

\begin{figure}
    \centering
    \includegraphics[width=1\textwidth]{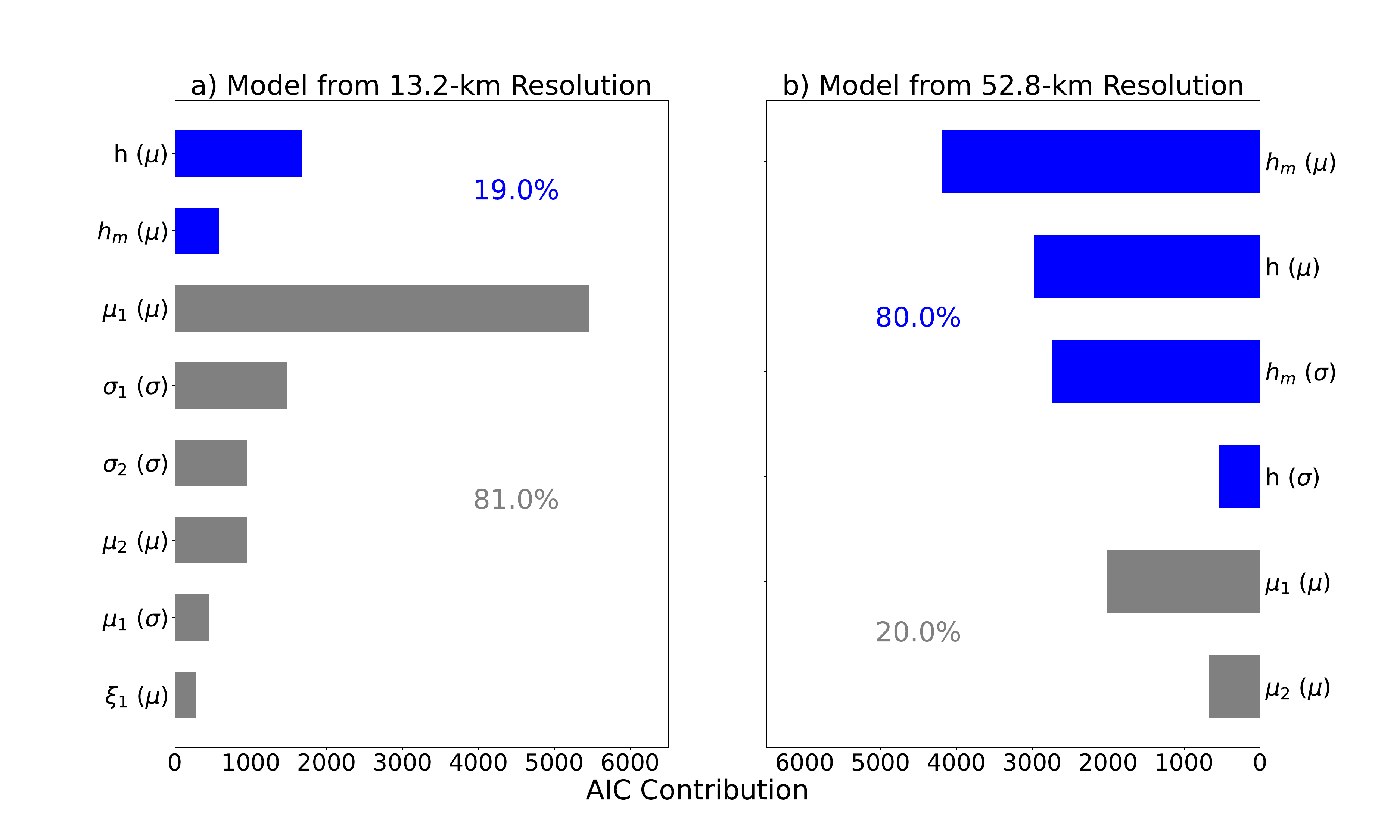} 
    \caption{Drop in Akaike Information Criterion (AIC) values, showing each feature’s explanatory power for the target GEV's location ($\mu$), scale ($\sigma$), and shape ($\xi$) parameters when super-resolving from 13.2\,km (a) and 52.8\,km (b). Coarse-resolution features (gray) become less informative than elevation statistics (blue) as the super-resolution factor increases from 6 (left) to 24 (right).} 
    \label{fig_aic} 
\end{figure}

Coarse-resolution features are dominant in this setup: they account for 81\% of the model's explanatory power, as determined by the AIC drop when each variable is removed (Figure~\ref{fig_aic}a). Among the coarse precipitation features, $\mu_1$ and $\sigma_1$ are most informative, while $\mu_2$ and $\sigma_2$ contribute less, suggesting that most of the information comes from the nearest coarse grid cell. The second-nearest block serves primarily as a correction, which is an expected and reassuring result. This configuration corresponds to a canonical super-resolution setting, where most of the predictive skill stems from the coarse field of interest rather than from external covariates.

\begin{figure}
    \centering
\includegraphics[width=1\textwidth]{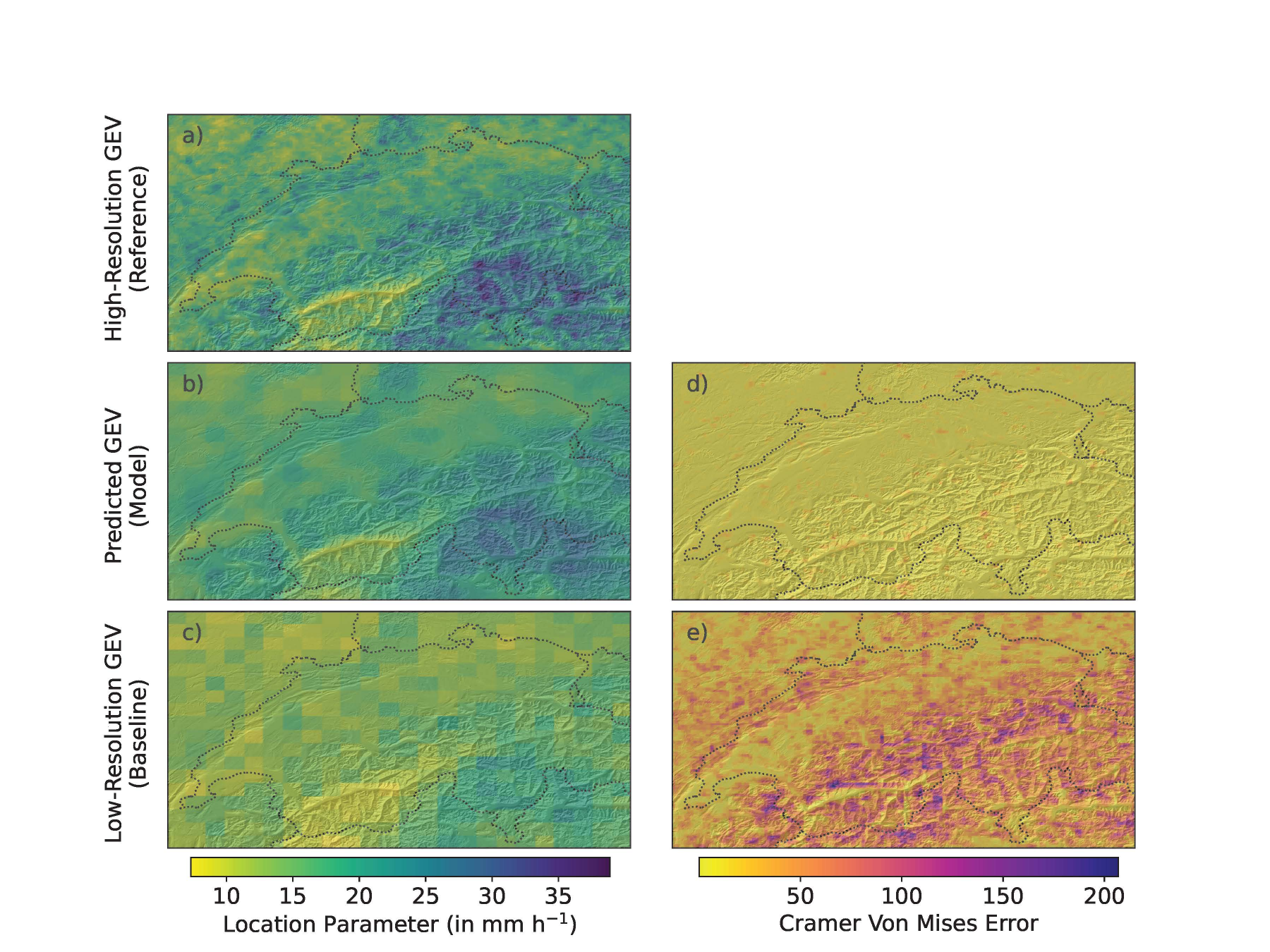} 
    \caption{Maps of GEV location parameters and errors using reference climate data. (Left) Location parameter values from the fine-resolution reference, VGAM prediction, and 13.2\,km-resolution baseline. (Right) Corresponding Cramér–von Mises errors for the model and the baseline. The VGAM improves spatial detail, particularly over complex terrain.}
    \label{fig:maps} 
\end{figure}

Figure~\ref{fig:maps} highlights the spatial improvements of the VGAM over the coarse-resolution GEV baseline. The right-hand maps show that VGAM predictions better capture fine-scale spatial variability than coarse-resolution baselines, especially over complex terrain such as the Alps. The model increases the location parameter in regions where extreme precipitation was previously underestimated, and captures spatial details more effectively, particularly in southern regions like Ticino. In contrast, models constrained to linear splines (VGLMs) are less flexible in capturing nonlinear relationships, which limits their ability to represent complex interactions between topography and precipitation extremes. The best-performing VGLM is defined as
\begin{equation}
\begin{cases}
    \mu(\bm{x}) = \beta_{\mu} + \beta_{\mu, h} x_h + \beta_{\mu, h_{\text{m}}} x_{h_{\text{m}}} + \beta_{\mu, \mu_1} x_{\mu_1} + \beta_{\mu, \mu_2} x_{\mu_2} + \beta_{\mu, \xi_1} x_{\xi_1} + \beta_{\mu, \xi_2} x_{\xi_2}, \\
    \log(\sigma(\bm{x})) = \beta_{\sigma} + \beta_{\sigma, \sigma_1} x_{\sigma_1} + \beta_{\sigma, \sigma_2} x_{\sigma_2}, \\
    \xi(\bm{x}) = \beta_{\xi}.
\end{cases}
\label{equa_VGLMmodel}
\end{equation}

The advantage of our trained models is supported by Table~\ref{tab:performance}, which reports the mean Cramér–von Mises error between the predicted and target distributions (the latter fitted to 2.2\,km-resolution precipitation). Both VGLM and VGAM outperform the coarse-resolution baseline, but the VGAM achieves lower errors across the training, validation, and test sets in the reference climate. We attribute this gain to the model’s greater ability to capture nonlinear relationships, underscored by VGAM’s consistent outperformance of VGLM.

\begin{table}
\caption{Comparison of VGAM and VGLM performance based on mean Cramér--von Mises errors across grid points. The first three columns show results for the training, validation, and test subsets within the climate used for model training (present or future). The last column reports performance over the entire domain in the climate not used for training, assessing spatial generalization across climates. VGAM consistently outperforms VGLM in the present (reference) climate.}
\centering
\resizebox{\textwidth}{!}{%
\begin{tabular}{c|c||c|c|c|c}
 & Model & Training Set & Validation Set & Test Set 1 & Test Set 2\tabularnewline 
 &  &  &  & \textit{Spatial} & \textit{Generalization}\tabularnewline 
Trained on  &  &  &  & \textit{Generalization} & \textit{Across Climates}\tabularnewline 

\hline
Present & Baseline & 40.92 & 33.68 & 38.44 & 17.60 \tabularnewline
13.2-km Resolution & VGAM & 3.81 & 4.05 & 3.61 & 5.18 \tabularnewline
(Upscaling factor: 6$\times$) & VGLM & 4.12 & 3.80 & 3.67 & 4.39 \tabularnewline

\hline
Present & Baseline & 106.99 & 90.23 & 100.13 & 51.08 \tabularnewline
26.4-km Resolution & VGAM & 4.52 & 5.61 & 5.51 & 10.10 \tabularnewline
(Upscaling factor: 12$\times$) & VGLM & 5.82 & 5.79 & 5.31 & 7.12    \tabularnewline

\hline
Present & Baseline & 189.05 & 185.52 & 182.23 & 107.12 \tabularnewline
52.8-km Resolution & VGAM & 5.50 & 9.45 & 6.05 & 12.57 \tabularnewline
(Upscaling factor: 24$\times$)& VGLM & 6.63 & 10.75 & 6.74 & 7.60 \tabularnewline

\hline
Future & Baseline & 18.23 & 15.37 & 17.29 & 39.30 \tabularnewline
13.2-km Resolution & VGAM & 3.02 & 3.55 & 3.55 & 6.26\tabularnewline
(Upscaling factor: 6$\times$) & VGLM & 3.13 & 3.52 & 3.55 & 6.42    \tabularnewline
\hline
Elevation & VGAM & 6.49 & 11.42 & 6.80  \tabularnewline
Features Only & VGLM & 9.41 & 15.94 & 7.91  \tabularnewline
\end{tabular}
}

\label{tab:performance}
\end{table}

\begin{figure}
    \centering
    \includegraphics[width=1\textwidth]{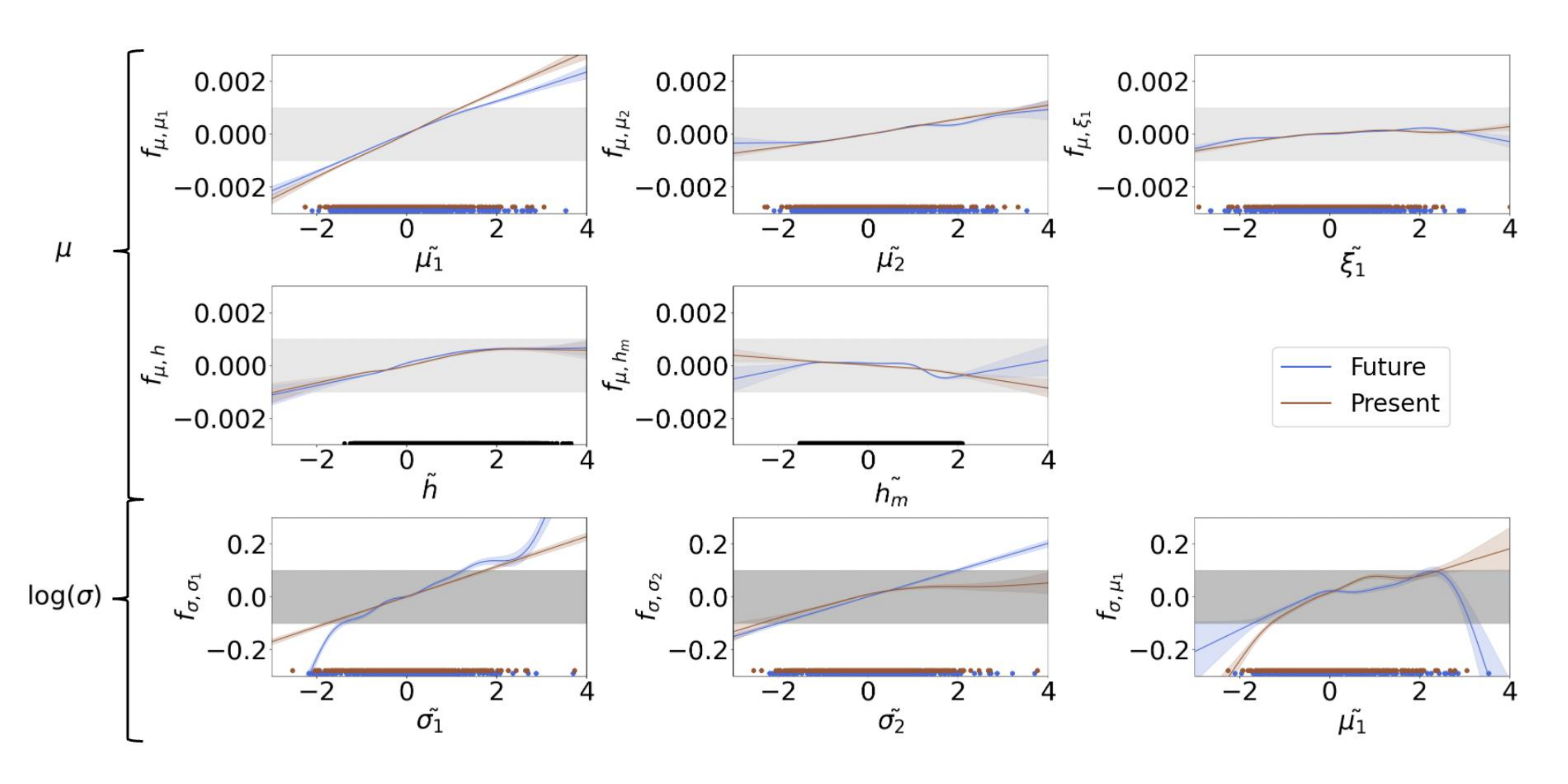} 
    \caption{Spline functions for the GEV parameters in~\eqref{equa_VGAMmodel}. The first five panels describe the additive components for the location parameter $\mu$, and the last three for $\log \sigma$. Each line shows the functions learned by models trained on present (brown) and future (blue) climate data, with 95\% confidence intervals overlaid and sample distributions shown on the x-axis.}
    \label{fig:splines} 
\end{figure}

One of the key advantages of VGAMs lies in their interpretability through smooth additive functions, which describe how each feature contributes to the predicted parameters of the target distribution. Figure~\ref{fig:splines} shows the learned splines for each feature in~\eqref{equa_VGAMmodel}, revealing meaningful relationships in the present climate (brown lines).

In the top two rows, which govern the location parameter $\mu$, the spline $f_{\mu, \mu_1}$ (top left) is monotonically increasing, indicating that higher coarse-scale location values predict higher fine-scale extremes. The spline $f_{\mu, \mu_2}$ (top middle) also increases, but with a lower slope, suggesting that the second-nearest block provides only a secondary correction. The spline $f_{\mu, h}$ (center left) rises with elevation before flattening at higher altitudes, consistent with orographic enhancement up to a threshold. The average elevation spline $f_{\mu, h_{\text{m}}}$ (center) refines this relationship by incorporating topographic context over a broader scale. In the bottom row, which models the logarithm of the scale parameter $\log(\sigma)$, similar patterns emerge. The spline $f_{\sigma, \sigma_1}$ (bottom left) increases nearly linearly, reinforcing the importance of coarse-scale scale values. The spline $f_{\sigma, \sigma_2}$ (bottom center) also increases but shows a threshold effect, indicating that secondary features affect the spread only beyond a certain magnitude. The function $f_{\sigma, \mu_1}$ (bottom right) shows a sharp, monotonic increase, revealing that regions with higher location values also exhibit greater variability.

\subsection{Generalization across Super-Resolution Factors}

To assess model generalization across super-resolution factors, we train VGAMs on 13.2\,km, 26.4\,km, and 52.8\,km input data, keeping the feature set fixed to that selected at 13.2\,km. At 26.4\,km, the model coefficients and splines remain similar, but performance degrades (Table~\ref{tab:performance}), indicating information loss from spatial coarsening (Figure~\ref{fig_blurring}). The 52.8\,km model diverges further: it selects fewer features and relies more on topography than on coarse GEV parameters. This shift marks a departure from a canonical super-resolution regime, where predictive skill stems from the coarse-resolution target field, toward a topographic downscaling setup. This is reflected in the 52.8\,km model equations:

\begin{equation}
\begin{cases}
    \mu(\bm{x}) = \beta_\mu + f_{\mu, h}(x_h) + f_{\mu, h_{\text{m}}}(x_{h_{\text{m}}}) + f_{\mu, h_{\text{s}}}(x_{h_{\text{s}}}) + f_{\mu, \mu_1}(x_{\mu_1}) + f_{\mu, \mu_2}(x_{\mu_2}), \\
    \log(\sigma(\bm{x})) = \beta_\sigma + f_{\sigma, h}(x_h) + f_{\sigma, h_{\text{m}}}(x_{h_{\text{m}}}), \\
    \xi(\bm{x}) = \beta_\xi,
\end{cases}
\label{equa_model48}
\end{equation}
where we remind the reader that $h_{\text{m}}$ and $h_{\text{s}}$ denote the mean and standard deviation of local elevation over the circular neighborhood of radius $R$.

Over the Swiss Alps, the 52.8\,km model using GEV features underperforms compared to an elevation-only model, reflecting the loss of predictive content in coarse precipitation fields. At such resolutions, alternative downscaling methods---such as perfect-prognosis approaches incorporating temperature or humidity, or deep learning architectures leveraging broader spatial contexts---may enhance performance.

\begin{figure}
    \centering
    \includegraphics[width=\textwidth]{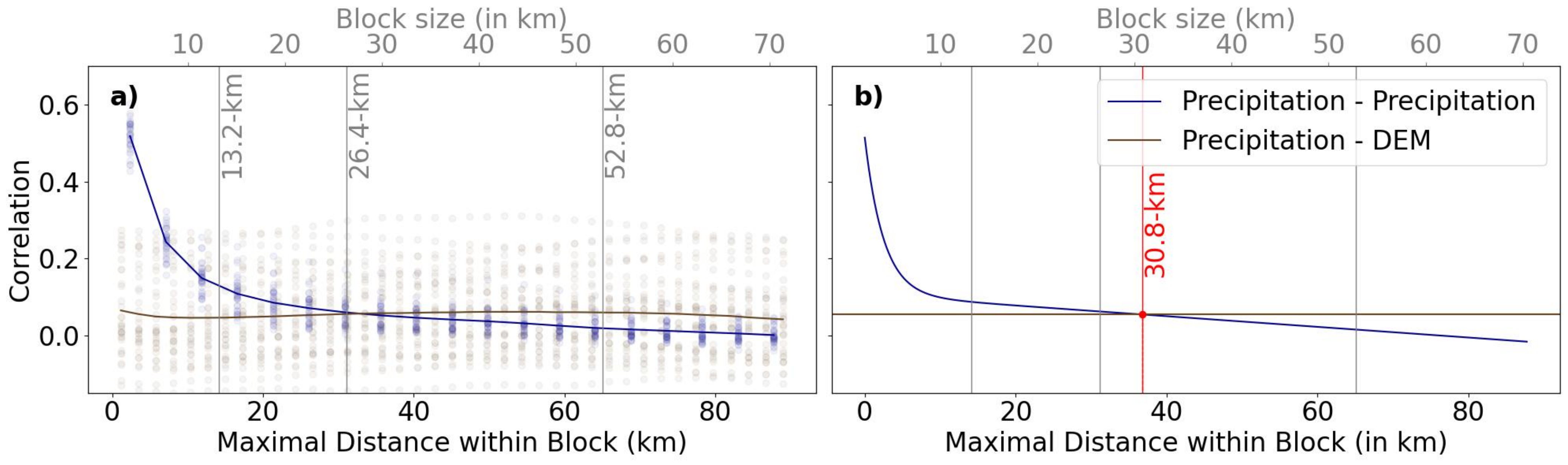}
    \caption{Spatial structure of precipitation and elevation fields. (a) Correlation function of precipitation (dotted blue) and cross-correlation with elevation (dotted brown) for each time stamp (month), with temporal means shown as solid lines (see ~\eqref{eq_spatial_correl_normalized} and~\eqref{eq_spatial_cross_correl_normalized}, respectively). Gray labels on the x-axis indicate block sizes matching the maximum radial distances for 13.2\,km, 26.4\,km, and 52.8\,km grids. (b) Fitted exponential decay (spatial correlation) and constant fit (spatial cross-correlation). Their intersection at a block size of 30.8\,km marks the limit beyond which spatial detail is insufficient for super-resolution.}
    \label{fig_corr}
\end{figure}

To better understand the observed performance drop, we apply the methodology described in Section~\ref{Subsec_Resol_BreakSR} to the the 2.2\,km precipitation and elevation fields. More precisely, we investigate the behavior of the spatial correlation and autocorrelation functions defined in~\eqref{eq_spatial_correl_normalized} and~\eqref{eq_spatial_cross_correl_normalized}. As illustrated in Figure~\ref{fig_corr}a, the spatial correlation of precipitation decreases approximately exponentially with distance, whereas the spatial cross-correlation between precipitation and elevation remains nearly constant. These behaviors can be modeled using the following functional forms, as shown in Figure~\ref{fig_corr}b:
\begin{equation}
    \rho_{PP}(r) = a \, \exp(-b r) + c r + d, \quad \rho_{PH}(r) \approx \text{const},
    \label{expo}
\end{equation}
where $r$ is radial distance, and $\rho_{PP}$ and $\rho_{PH}$ are the spatial correlation and cross-correlation functions, respectively. Interpreting $r$ as the maximum radial distance between two blocks of side length $2.2$\,km within a block of side length $\delta x$ (maximum distance appearing in~\eqref{Eq_resolution_break_SR}), the curves intersect at $r \approx 38$\,km, or $\delta x \approx 30.8$\,km. Beyond this, elevation becomes more informative than precipitation, marking the end of the super-resolution regime.

This provides a heuristic threshold: once block size exceeds 30.8\,km, coarse precipitation loses predictive advantage, and topographic features dominate. In flatter terrain, the heuristic could be adapted using alternative covariates or by identifying a correlation cutoff (e.g., 0.1) below which super-resolution is no longer effective; see details in Section~\ref{Subsec_Resol_BreakSR}.

\subsection{Understanding Robustness to Climate Change}

\begin{figure}
    \centering
    \includegraphics[width=1\textwidth]{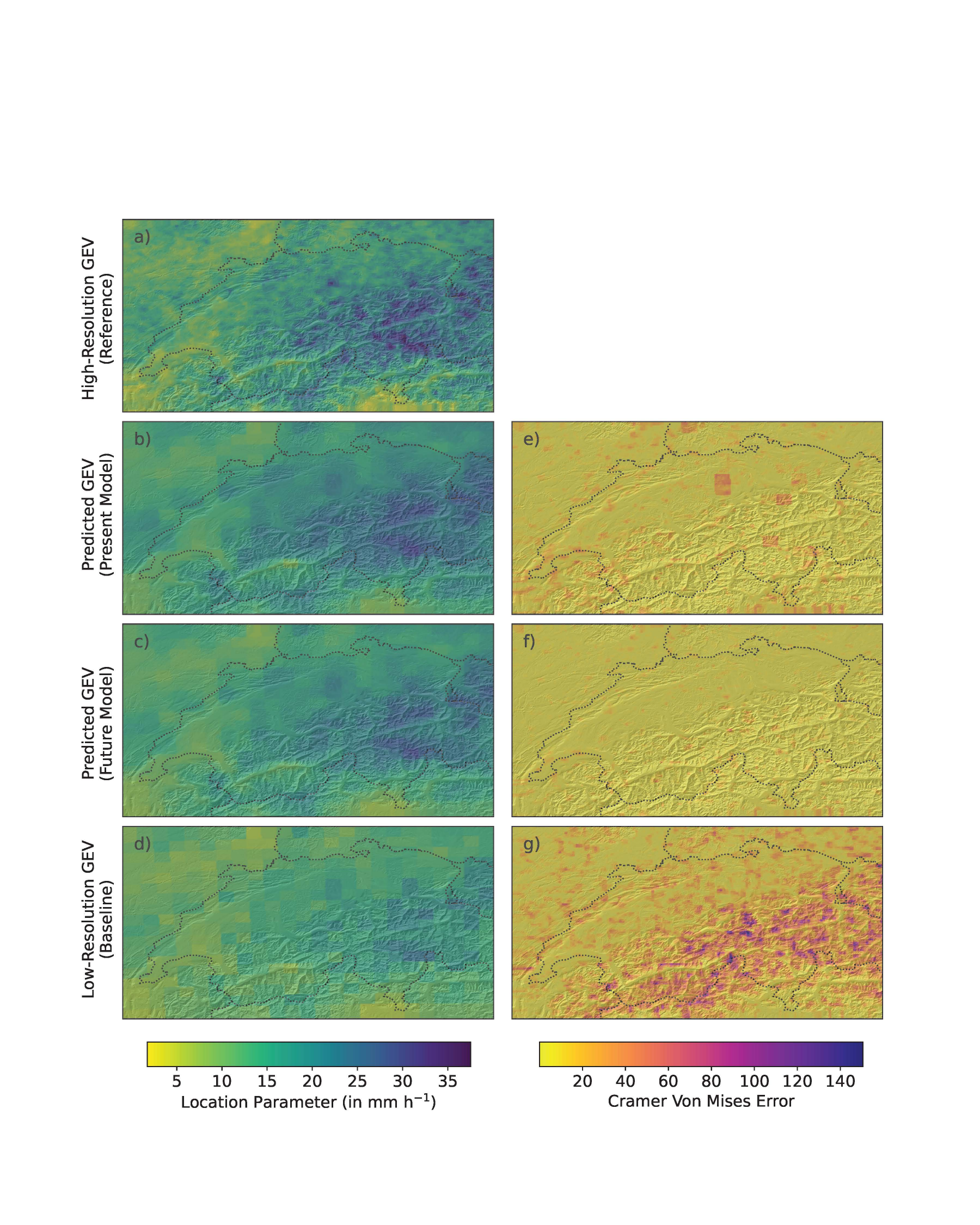} 
    \caption{Maps of GEV location parameters and Cramér–von Mises errors using data from the pseudo-global warming simulation. (Left, from top to bottom) Location parameter values from the fine-resolution reference, VGAM predictions trained on present and warmed climates, and the 13.2\,km-resolution baseline. (Right) Corresponding Cramér–von Mises errors for each model.}
    \label{fig:maps_future}
\end{figure}

Finally, we use our distribution super-resolution framework to assess challenges in generalization/robustness across climates. Climate change is treated here as a domain shift in model features, affecting the location, scale, and shape parameters of the coarse-resolution GEV distributions. As model errors increase in the future climate, where precipitation extremes intensify, we evaluate robustness by comparing the performance on future data of reference models (trained on present-day data) to that of models trained directly on future climate data (see Figure~\ref{fig:maps_future}).

This comparison enables an assessment of both spatial generalization and generalization across climates. When recalibrated on future data, the VGAM retains qualitatively similar splines as shown in Figure~\ref{fig:splines}, but the intercepts for both the scale and shape parameters increase, with the shape parameter rising from 0.19 to 0.25.

To diagnose the source of generalization errors, we analyze the robustness gap introduced in Section~\ref{Subsec_GeneralizabilityRobustnessCC}, defined via the pinball loss, which quantifies the discrepancy between predicted quantiles for future climates when using models trained in present and future climates. In the case of the GEV distribution, we decompose the quantile gap $\Delta_{\alpha}$ (difference between the quantiles for future climate, computed from a model trained on present-day data and a model trained on future data) at a specific level $\alpha \in (0, 1)$ using a first-order Taylor expansion:
\begin{equation}
\Delta_\alpha = \left( \frac{\partial q_\alpha}{\partial \mu} \right) (\mu^\text{P} - \mu^\text{F}) + \left( \frac{\partial q_\alpha}{\partial \sigma} \right) (\sigma^\text{P}-\sigma^\text{F}) + \left( \frac{\partial q_\alpha}{\partial \xi} \right) (\xi^\text{P} - \xi^\text{F}) + R_{\Delta_\alpha},
\end{equation}
where $ R_{\Delta_\alpha}$ is the residual, and the partial derivatives of the GEV quantile function expressed in~(S3) of the SM are given by
\begin{equation}
\frac{\partial q_\alpha}{\partial \mu} = 1, \quad 
\frac{\partial q_\alpha}{\partial \sigma} = \frac{(-\log \alpha)^{-\xi} - 1}{\xi}, \quad 
\frac{\partial q_\alpha}{\partial \xi} = \frac{(-\log \alpha)^{-\xi} \left(-\xi \log(-\log \alpha) - 1\right) + 1}{\xi^2};
\end{equation}
for the detailed derivation of these terms, see~Sections C.2 and C.3 of the SM. 

By incorporating this decomposition into~\eqref{Eq_1stExpressions_RG}, we obtain the decomposed form of the pointwise robustness gap $G$ in the GEV case:
\begin{equation}
\label{Eq_Decompos_PointwiseRobustnessGapGEV}
G = \text{T}_\mu + \text{T}_\sigma + \text{T}_\xi + R + f(\varepsilon_{\text{F}, \alpha}, \Delta_{\alpha}),
\end{equation}
where
\begin{equation}
\begin{cases}
    \text{T}_\mu =  \frac{\partial Q_\alpha}{\partial \mu} (\mu^\text{P} - \mu^\text{F}) \left( \mathbb{I}_{\{\varepsilon_{\text{F}, \alpha} > -\Delta\}} - \alpha \right), \\
    \text{T}_\sigma =  \frac{\partial Q_\alpha}{\partial \sigma} (\sigma^\text{P} - \sigma^F) \left( \mathbb{I}_{\{\varepsilon_{\text{F}, \alpha} > -\Delta_\alpha\}} - \alpha \right), \\
    \text{T}_\xi =  \frac{\partial Q_\alpha}{\partial \xi} (\xi^\text{P} - \xi^\text{F}) \left( \mathbb{I}_{\{\varepsilon_{\text{F}, \alpha} > -\Delta_\alpha\}} - \alpha \right), \\
    R =   R_{\Delta_\alpha} \left( \mathbb{I}_{\{\varepsilon_{\text{F}, \alpha} > -\Delta_\alpha\}} - \alpha \right), \\
    f(\varepsilon_{\text{F}, \alpha}, \Delta_\alpha) = \varepsilon_{\text{F}, \alpha} \left( \mathbb{I}_{\{\varepsilon_{\text{F}, \alpha} > -\Delta_\alpha\}} - \mathbb{I}_{\{\varepsilon_{\text{F}, \alpha} > 0\}} \right).
\end{cases}
\label{decomposition_RG}
\end{equation}

\begin{figure}
    \centering
    \includegraphics[width=1\textwidth]{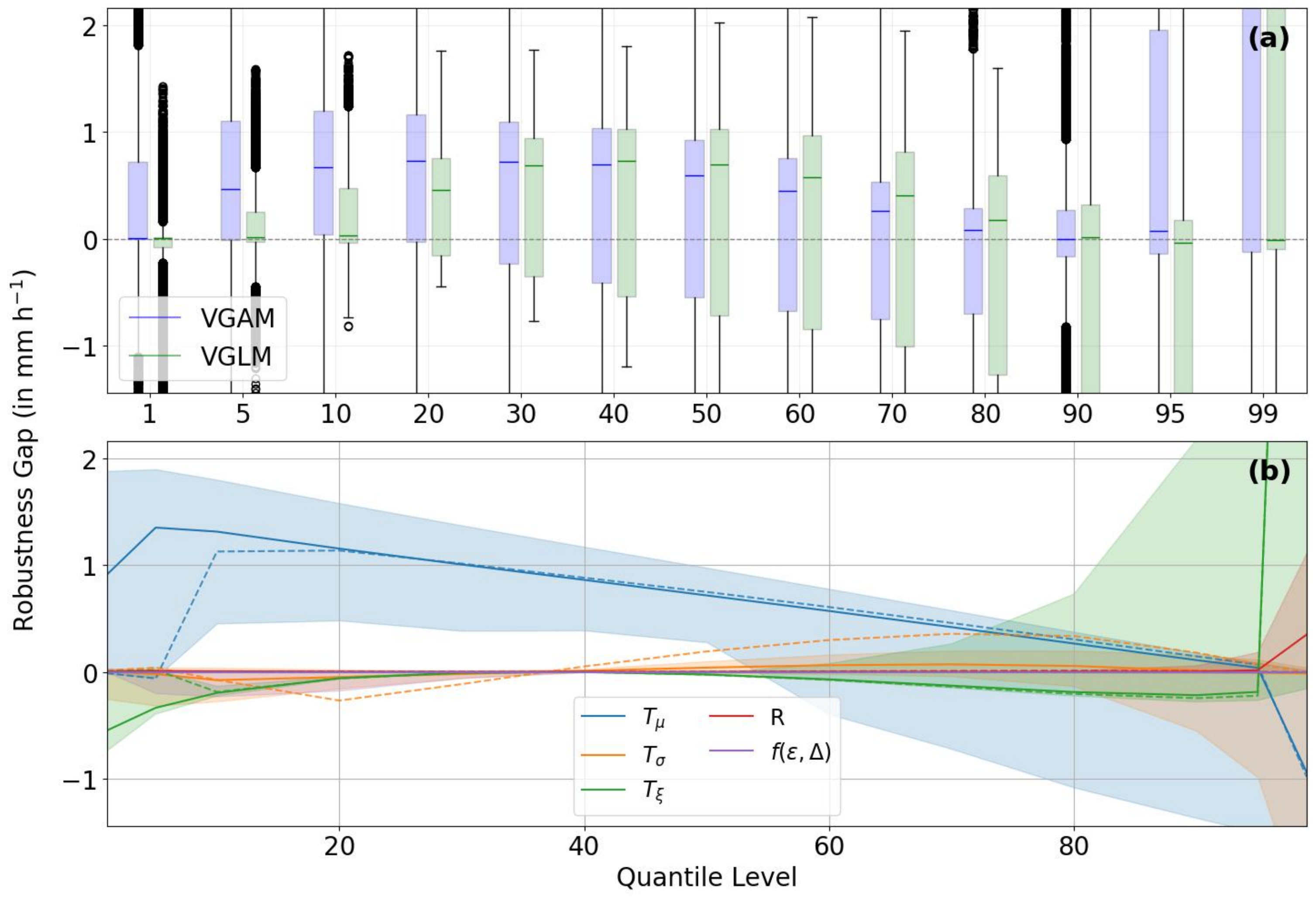}
    \caption{(a) Boxplots of the pointwise robustness gaps across quantile levels for the VGAM (blue) and VGLM (green) models. Each boxplot has been built using the pointwise robustness gaps of all grid points. (b) Contribution of individual terms in the pointwise robustness gap decomposition (see~\eqref{Eq_Decompos_PointwiseRobustnessGapGEV}). Solid lines indicate the median, with shaded areas representing the interquartile range. Dashed lines denote the VGLM median baseline for comparison. Median and quartiles are computed over all grid points.}
    \label{fig:robustness_gap}
\end{figure}

Panel~(a) of Figure~\ref{fig:robustness_gap} shows very large pointwise robustness gaps at high quantile levels (above the 90\textsuperscript{th} percentile), revealing clear limitations in the model's ability to generalize under climate change. Panel~(b) indicates that this degradation is mainly driven by the scale and shape parameters, highlighting the importance of accounting for changes in dispersion and tail behavior. In particular, the assumption of a spatially constant shape parameter in the VGAM strongly limits the model’s capacity to capture modifications in the distribution tails under future conditions.

\begin{figure}
    \centering
    \includegraphics[width=1\textwidth]{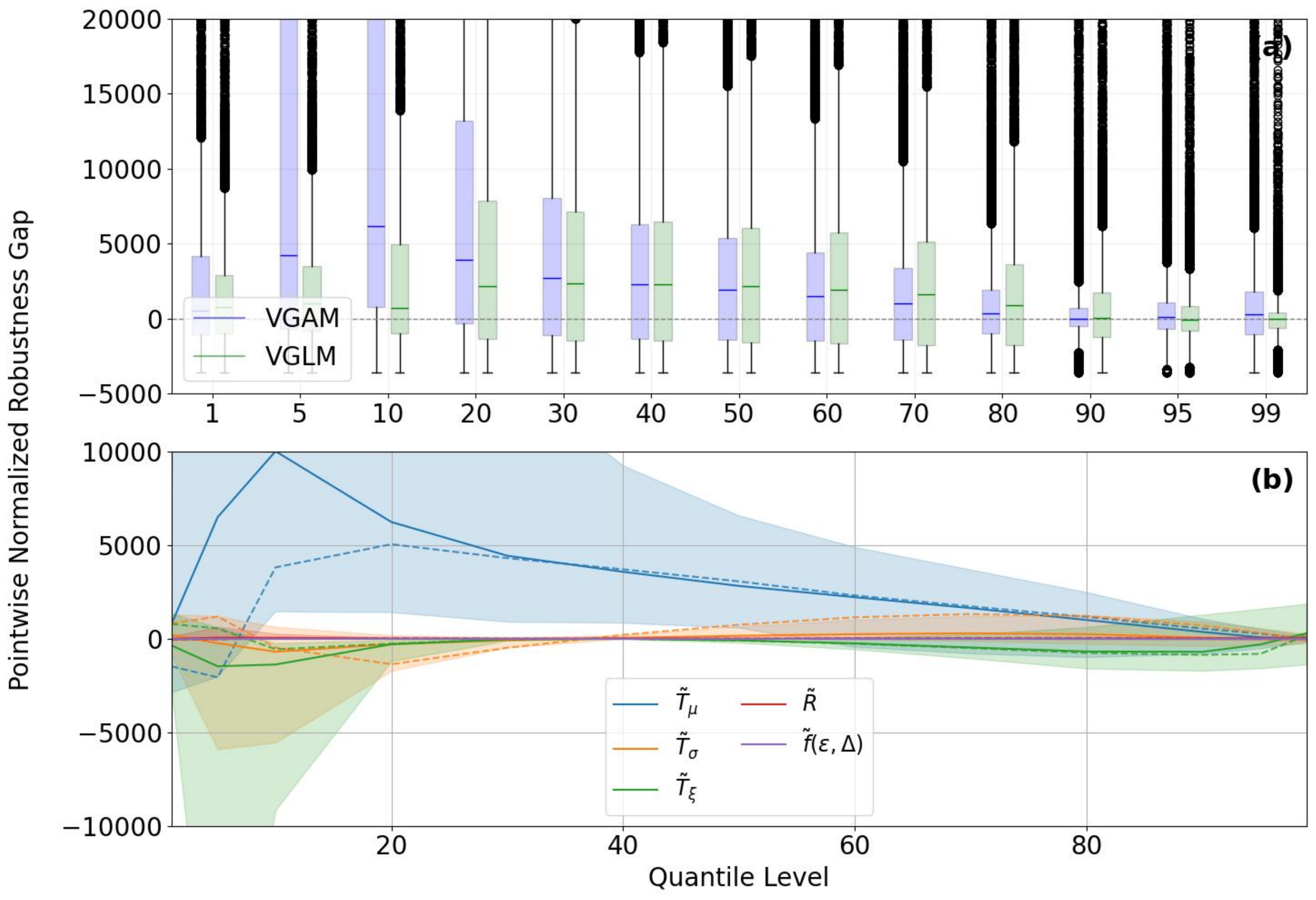}
    \caption{Same as Figure~\ref{fig:robustness_gap}, with all values---pointwise robustness gaps in (a) and contributions in (b)---normalized by $\ell_\alpha\left(q_{\text{F}, \alpha}, q^{\text{F}}_{\text{F}, \alpha}\right)$. The normalization of the contributions in~\eqref{Eq_Decompos_PointwiseRobustnessGapGEV} is indicated by adding a tilde symbol.}
    \label{fig:robustness_gap_normalized}
\end{figure}

This interpretation must, however, be nuanced. In the normalized robustness gap (Figure~\ref{fig:robustness_gap_normalized}), the sharp increase at high quantiles is clearly attenuated, indicating that the large absolute gaps occur in a regime where the model trained on future already exhibits high losses. This reflects the intrinsic difficulty of predicting very rare extremes rather than solely a lack of model transferability.


At lower quantiles, Figure~\ref{fig:robustness_gap}(a) shows better generalization of the VGLM, with consistently smaller absolute pointwise robustness gaps. In contrast, Figure~\ref{fig:robustness_gap_normalized}(a) reveals relatively large normalized gaps at the 5\textsuperscript{th} and 10\textsuperscript{th} percentiles, mainly due to the very low losses of the future model in this regime. Across both figures, Panel~(b) consistently shows that the location parameter term $T_\mu$ dominates the robustness gap, while scale and shape contributions remain negligible, underscoring the central role of the location parameter in governing robustness at low quantiles.

\section{Summary and Conclusions} 
\label{sec_conclusion}

\subsection{Summary}
\begin{figure}
    \centering    
    \includegraphics[width=\textwidth]{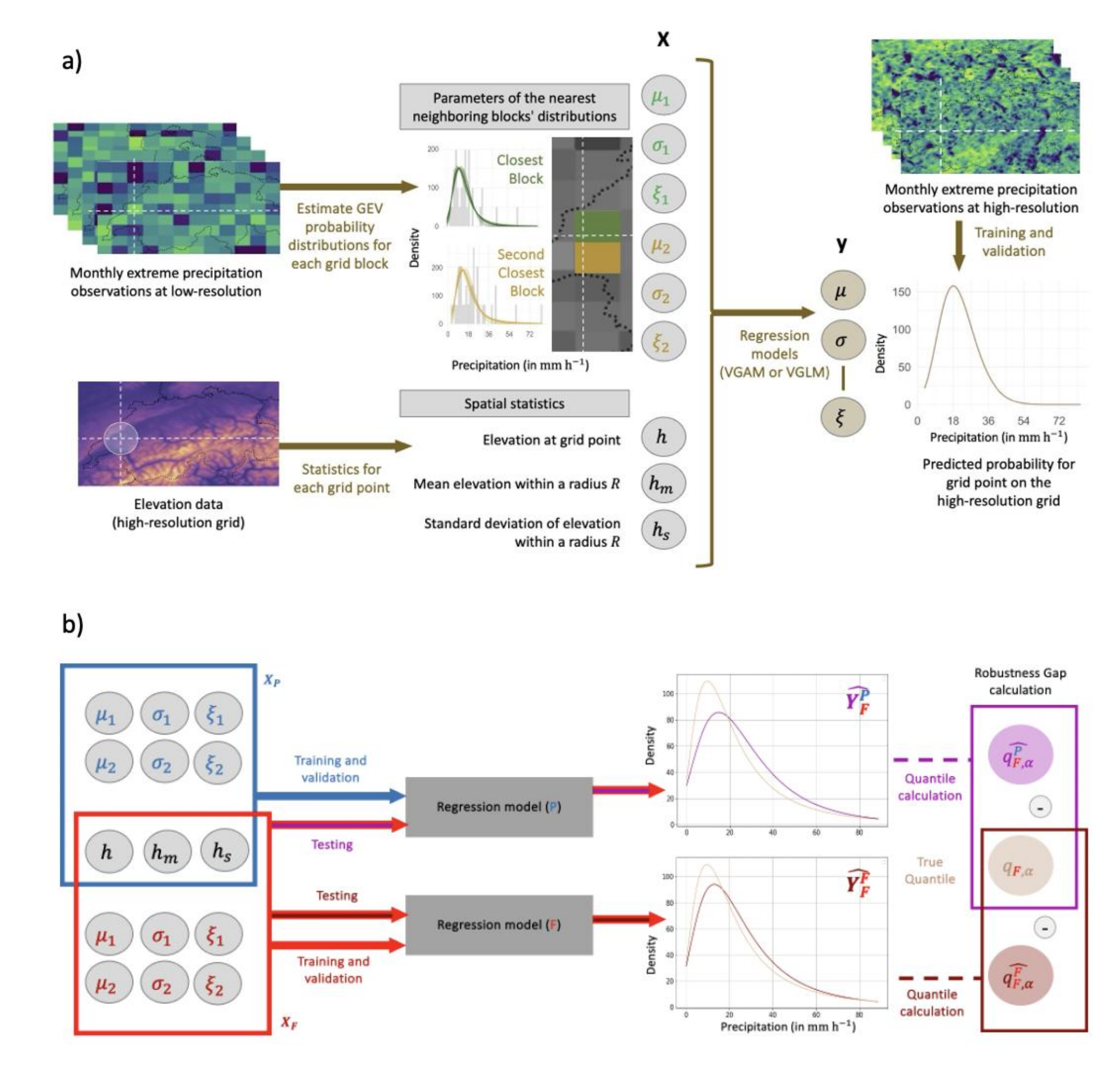} 
    \caption{(a) Our interpretable super-resolution model predicts GEV probability distributions of extreme precipitation at 2\,km resolution by combining coarse-resolution precipitation data from nearby locations. Specifically, it predicts the location ($\mu$) and scale ($\sigma$) parameters as functions of the nearest neighbors' distribution parameters ($\mu_{1/2}$, $\sigma_{1/2}$, $\xi_{1/2}$) and topographic spatial statistics ($h$, $h_{\text{m}}$, $h_{\text{s}}$). (b) We quantitatively evaluate the generalizability of our framework using a pseudo-reality setup in which models trained in the historical (blue) and projected (red) climates are compared on future data. A robustness gap is computed for each precipitation quantile $q$.}
    \label{fig_intro} 
\end{figure}

As illustrated in Figure \ref{fig_intro}, we introduced a framework for super-resolving distributions within interpretable statistical models, focusing on risk-relevant quantities such as high quantiles (return levels). This approach is particularly valuable for impact modeling, as it enables the estimation of risk measures at resolutions suited to downstream applications. We developed a method to identify the resolution scale at which super-resolution begins to fail and proposed the novel concept of a robustness gap, which we analyzed in the context of quantile estimation using the pinball loss. Applied in a pseudo-reality experiment over Switzerland, our framework showed that the quantile-wise robustness gap is an effective diagnostic for evaluating how well models trained on present-day data generalize to warmer climates. Leveraging a GEV distribution parameterized via VGAMs, we could pinpoint which model components--parameters and splines--contribute to robustness failures, offering insights into model limitations and guiding future improvements. Overall, the proposed methodology enables a tractable decomposition of spatial generalization and robustness across climates. Our framework is distribution-agnostic and extends beyond precipitation to variables with tractable distributions (e.g., temperature extrema, wind speeds/gusts, and river discharge). By contrast, choices such as using a GEV distribution for block-maxima, elevation-based covariates, and mean pooling for coarsening are precipitation/region-specific and may be adapted per variable and application.

\subsection{Outlook and Conclusions}
Several directions could further improve this framework. First, our use of mean pooling to approximate coarse-resolution fields is a first-order simplification; future work could explore alternative filters (e.g., Gaussian) or incorporate temporal and physical biases through explicit bias correction. Second, we assumed a constant shape parameter for the GEV distribution, which may not reflect how extremes respond to climate shifts. Using the $r$-largest values approach could stabilize tail estimates and support state-dependent shape parameter modeling, helping address an open question: which covariates most strongly influence the shape parameter $\xi$? Third, since the original pseudo-global warming simulation used here \citep{Hentgen2019data}, the pseudo-global warming method has been refined \citep{Brogli_2023,Heim_2023}, and future work should update the analysis using the latest available simulations. Finally, tailored versions of the GEV, such as the blended GEV \citep{vandeskog2022modelling}, may improve the stability and realism of tail estimates. 

Overall, our results suggest that combining probabilistic super-resolution with quantifiable generalization/robustness diagnostics provides a principled framework for modeling extremes under distributional transformations, with potential relevance to other fields that study rare events in non-stationary systems.


\section*{Acknowledgments}

We thank Marine Berthier and Leo Micollet for preliminary work that laid useful foundations for this study, Linda Mhalla and Shivanshi Asthana for feedback that improved our manuscript, and Laureline Hentgen, Nikolina Ban and Christoph Sch\"{a}r, for performing the kilomter scale PGW simulations. These simulations were conducted on the Piz Daint supercomputer of the Swiss National Supercomputing Centre (CSCS) under project ID pr144. TB acknowledges support from the Swiss National Science Foundation (SNSF) under Grant No. 10001754 (``RobustSR'' project). EK would like to thank the Expertise Center for Climate Extremes (ECCE) at UNIL for financial support.  

\bibliographystyle{elsarticle-harv} 
\bibliography{sample}

\end{document}


\renewcommand{\theequation}{S\arabic{equation}}
\renewcommand{\thefigure}{S\arabic{figure}}

\date{}

\maketitle

\appendix

This Supplementary Material provides technical derivations that support Sections~3, ~4 and~5 of the manuscript. It includes the proof of an equation for the pointwise robustness gap, details the estimation of the parameters of the Generalized Extreme-Value (GEV) distribution, as well as those of a Vector Generalized Additive Model (VGAM) when used to model the GEV parameters. Additionally, it includes the expression of the GEV quantile function and its partial derivatives with respect to the location, scale, and shape parameters. Finally, it examines the potential non-stationarity in the time series of precipitation.

\section{Proof of the Formula for the Pointwise Robustness Gap (Equation~(18))}

The pointwise robustness gap defined in (18) can be expressed as
\begin{align*}
\label{Eq_1stExpressions_RG}
\mathrm{PRG} & 
\overset{\mathrm{def}}{=}
\ell_\alpha \left(q_{\text{F}, \alpha}, q^{\text{P}}_{\text{F}, \alpha} \right) - \ell_\alpha\left(q_{\text{F}, \alpha}, q^{\text{F}}_{\text{F}, \alpha}\right) \notag
 \\&= \left(\alpha - \mathbb{I}_{\{q_{\text{F}, \alpha} < q^{\text{P}}_{\text{F}, \alpha}\}}\right)\left(q_{\text{F}, \alpha} - q^{\text{P}}_{\text{F}, \alpha}\right) - \left(\alpha - \mathbb{I}_{\{q_{\text{F}, \alpha} < q^{\text{F}}_{\text{F}, \alpha}\}}\right)\left(q_{\text{F}, \alpha} - q^{\text{F}}_{\text{F}, \alpha}\right) \notag \\
&= -\alpha\left(q^{\text{P}}_{\text{F}, \alpha} - q^{\text{F}}_{\text{F}, \alpha}\right) - \mathbb{I}_{\{q_{\text{F}, \alpha} < q^{\text{P}}_{\text{F}, \alpha}\}}\left(q_{\text{F}, \alpha} - q^{\text{P}}_{\text{F}, \alpha}\right) + \mathbb{I}_{\{q_{\text{F}, \alpha} < q^{\text{F}}_{\text{F}, \alpha}\}}\left(q_{\text{F}, \alpha} - q^{\text{F}}_{\text{F}, \alpha}\right) \notag \\ 
&= -\alpha \Delta_{\alpha} + \mathbb{I}_{\{\varepsilon_{\text{F}, \alpha} + \Delta_{\alpha} > 0\}}\left( \varepsilon_{\text{F}, \alpha} + \Delta_{\alpha} \right) - \mathbb{I}_{\{\varepsilon_{\text{F}, \alpha} > 0\}} \varepsilon_{\text{F}, \alpha} \notag \\
&= \Delta_{\alpha} \left( \mathbb{I}_{\{\varepsilon_{\text{F}, \alpha} > -\Delta_{\alpha}\}} - \alpha \right) + \varepsilon_{\text{F}, \alpha} \left( \mathbb{I}_{\{\varepsilon_{\text{F}, \alpha} > -\Delta_{\alpha}\}} - \mathbb{I}_{\{\varepsilon_{\text{F}, \alpha} > 0\}} \right).
\end{align*}
This completes the derivation of (18).

\section{Estimation of GEV Parameters}
\label{App_Loglik}

\subsection{Classical Setting}

Assume that we have extracted $m$ block-maxima denoted by $M_1,\dots,M_m$. Under the independence and identical distribution assumption, the log‐likelihood of the GEV parameters (location $\mu$, scale $\sigma$, shape $\xi$) is
\begin{equation}
   \label{eq:gevl} 
\ell(\mu,\sigma,\xi)
=
\sum_{i=1}^m
\left \{
- \log\sigma 
- \Bigl(1+\tfrac{1}{\xi}\Bigr)\log\left (1+\xi\,\tfrac{M_i-\mu}{\sigma}\right)
- \left ( 1+\xi\,\tfrac{M_i-\mu}{\sigma}\right)^{-1/\xi}
\right \},
\end{equation}
and the maximum‐likelihood estimates $\hat\mu,\hat\sigma,\hat\xi$ are obtained by numerically maximizing $\ell$.

\subsection{VGAM Setting}
\label{Subsec_Estimation_VGAM_GEV}

Let us recall that, in the VGAM framework, the GEV parameters are expressed as
$$
\begin{aligned}
\mu_i &= \eta_1(\bm{x}_i)
       = \beta_{\mu} + \sum_{k=1}^p f_{\mu,k}(x_{ik})\,,\\
\log \sigma_i &= \eta_2(\bm{x}_i)
             = \beta_{\sigma} + \sum_{k=1}^p f_{\sigma,k}(x_{ik})\,,\\
\xi_i &= \eta_3(\bm{x}_i)
        = \beta_{\xi} + \sum_{k=1}^p f_{\xi,k}(x_{ik})\,,
\end{aligned}
$$
where $\beta_{\mu}$, $\beta_{\sigma }$ and  $\beta_{\xi}$ are the intercepts for the parameters $\mu$, $\sigma$ and $\xi$; $f_{\mu,k}(\cdot)$, $f_{\sigma,k}(\cdot)$ and $f_{\xi,k}(\cdot)$ are possibly smooth functions of the $k$-th covariate.

We estimate all unknowns by maximizing a penalized log-likelihood over the parameter vector
$$
\bm{\theta} = \{\beta_{\mu}, \beta_{\sigma}, \beta_{\xi}, \bm{b}_{\mu,k},\bm{b}_{\sigma,k},\bm{b}_{\xi,k} :\;k=1,\dots,p\},
$$
where $\bm{b}_{\mu,k}$ (respectively $\bm{b}_{\sigma,k}$ and $\bm{b}_{\xi,k}$) collects the basis coefficients for $f_{\mu,k}$ (respectively $f_{\sigma,k}$ and $f_{\xi,k}$).  Writing $\ell(\bm{\theta})$ for the GEV log-likelihood  \eqref{eq:gevl},
the penalized objective is
\begin{equation}
\label{Eq_Penalized_Loglik}
\ell_p(\bm{\theta}) 
= \ell(\bm{\theta}) 
  \;-\; \sum_{k=1}^p \tfrac{1}{2}\,\lambda_{\mu,k}\, \bm{b}_{\mu,k}^{\top} \bm{S}\,\bm{b}_{\mu,k}
    \;-\;  
   \sum_{k=1}^p \tfrac{1}{2}\,\lambda_{\sigma,k}\, \bm{b}_{\sigma,k}^{\top} \bm{S}\,\bm{b}_{\sigma,k}
    \;-\;   
      \sum_{k=1}^p \tfrac{1}{2}\,\lambda_{\xi,k}\, \bm{b}_{\xi,k}^{\top} \bm{S}\,\bm{b}_{\xi,k},
\end{equation}
where $\bm{S}$ is a $m\times m$ penalty matrix (e.g.,\ second-derivative penalty for splines), $m$ being the number of basis functions used for that smoothing and $\lambda_{\mu,k}\ge0$ (respectively $\lambda_{\sigma,k}\ge0$ and $\lambda_{\xi,k}\ge0$) is the smoothing parameter controlling the wiggliness of $f_{\mu,k}$ (respectively $f_{\sigma,k}$ and $f_{\xi,k}$).

\medskip

\noindent\textbf{Algorithmic details:}
\begin{itemize}
  \item We solve for $\bm{\theta}$ by a penalized Fisher scoring (or quasi‐Newton) algorithm, iterating until convergence.
  \item The smoothing parameters $\lambda_{\mu,k}$, $\lambda_{\sigma,k}$ and $\lambda_{\xi,k}$ are chosen automatically, e.g.\ by minimizing a generalized cross-validation (GCV) criterion or by approximate restricted maximum likelihood (REML).
  \item The standard errors for $\eta_j(\bm{x})$  ($j=1,2,3$) and  for any function of the GEV parameters are obtained from the inverse penalized Fisher information.
\end{itemize}

\section{Partial Derivatives of the GEV Quantile Function}
\label{Sec_GEVQuantileFunction}

\subsection{GEV Quantile Function}

The quantile function at level $\alpha \in (0,1)$ of the GEV distribution with location, scale and shape parameters $\mu$, $\sigma$ and $\xi$, is
\begin{equation}
\label{Eq_QuantileFunction_GEV}
q_\alpha = 
\begin{cases}
\mu + \dfrac{\sigma}{\xi} \left[ (-\log \alpha)^{-\xi} - 1 \right], & \text{for } \xi \neq 0, \\[10pt]
\mu - \sigma \log(-\log \alpha), & \text{for } \xi = 0.
\end{cases}
\end{equation}

In this appendix, we focus on the case $\xi \neq 0$. While $\xi$ may approach zero in some regions, the GEV quantile function for $\xi \neq 0$ continuously converges to the $\xi = 0$ (Gumbel) case as $\xi \to 0$. Therefore, using the \(\xi \neq 0\) expression provides a unified and consistent formulation across all locations without requiring a separate derivation. In practice, this approximation is accurate and numerically stable even for small values of \(|\xi|\) \citep{GEV}.

\subsection{Partial Derivative with respect to \(\mu\) and \(\sigma\)}

For the partial derivatives with respect to \(\mu\) and \(\sigma\) we immediately obtain
\begin{equation}
\frac{\partial q_\alpha}{\partial \mu} = 1, \quad \frac{\partial q_\alpha}{\partial \sigma} = \frac{1}{\xi} \left[ (-\log \alpha)^{-\xi} - 1 \right].
\end{equation}

\subsection{Partial Derivative with respect to $\xi$}

Let $A = (-\log \alpha)^{-\xi}$ for brevity. Then we have
\begin{equation}
\frac{\partial q_\alpha}{\partial \xi} = \frac{\partial}{\partial \xi} \left( \frac{\sigma}{\xi} (A - 1) \right).
\end{equation}
Applying the product and chain rules, we get
\begin{equation}
\frac{\partial q_\alpha}{\partial \xi} = \sigma \left( \frac{-1}{\xi^2} (A - 1) + \frac{1}{\xi}  \frac{\partial A}{\partial \xi} \right),
\end{equation}
where
\begin{equation}
\begin{aligned}
\frac{\partial A}{\partial \xi} 
= \frac{\partial}{\partial \xi} \left[ (-\log \alpha)^{-\xi} \right] 
= \frac{\partial}{\partial \xi} \left[ e^{-\xi \log(-\log \alpha)} \right] 
&= e^{-\xi \log(-\log \alpha)} \left[ - \log(-\log \alpha) \right] \\&
= (-\log \alpha)^{-\xi} \left[ - \log(-\log \alpha) \right].
\end{aligned}
\end{equation}
Putting everything together yields
\begin{equation}
\begin{aligned}
\frac{\partial q_\alpha}{\partial \xi} 
&= \sigma \left( -\frac{1}{\xi^2} \left[ (-\log \alpha)^{-\xi} - 1 \right] + \frac{1}{\xi} \frac{\partial}{\partial \xi} \left[ (-\log \alpha)^{-\xi} \right] \right) \\
&= \sigma \left( -\frac{1}{\xi^2} \left[ (-\log \alpha)^{-\xi} - 1 \right] + \frac{1}{\xi} (-\log \alpha)^{-\xi} \left[ -\log(-\log \alpha) \right] \right) \\
&= \frac{\sigma}{\xi^2} \left( -(-\log \alpha)^{-\xi} + \xi (-\log \alpha)^{-\xi}  \left[ -\log(-\log \alpha) \right] + 1 \right) \\
&= \frac{\sigma}{\xi^2} \left[ (-\log \alpha)^{-\xi} \left( -\xi \log(-\log \alpha) - 1 \right) + 1 \right].
\end{aligned}
\end{equation}

\section{Temporal Stationarity of Extreme Precipitation Values}
\label{sec:appendix_temporal}

This appendix aims to justify the exclusion of a temporal trend term from our super-resolution model. Both present-day and future simulations cover an 11-year period, with three block maxima per year extracted at each grid point.


\begin{figure}
    \centering
    \includegraphics[width=1\textwidth]{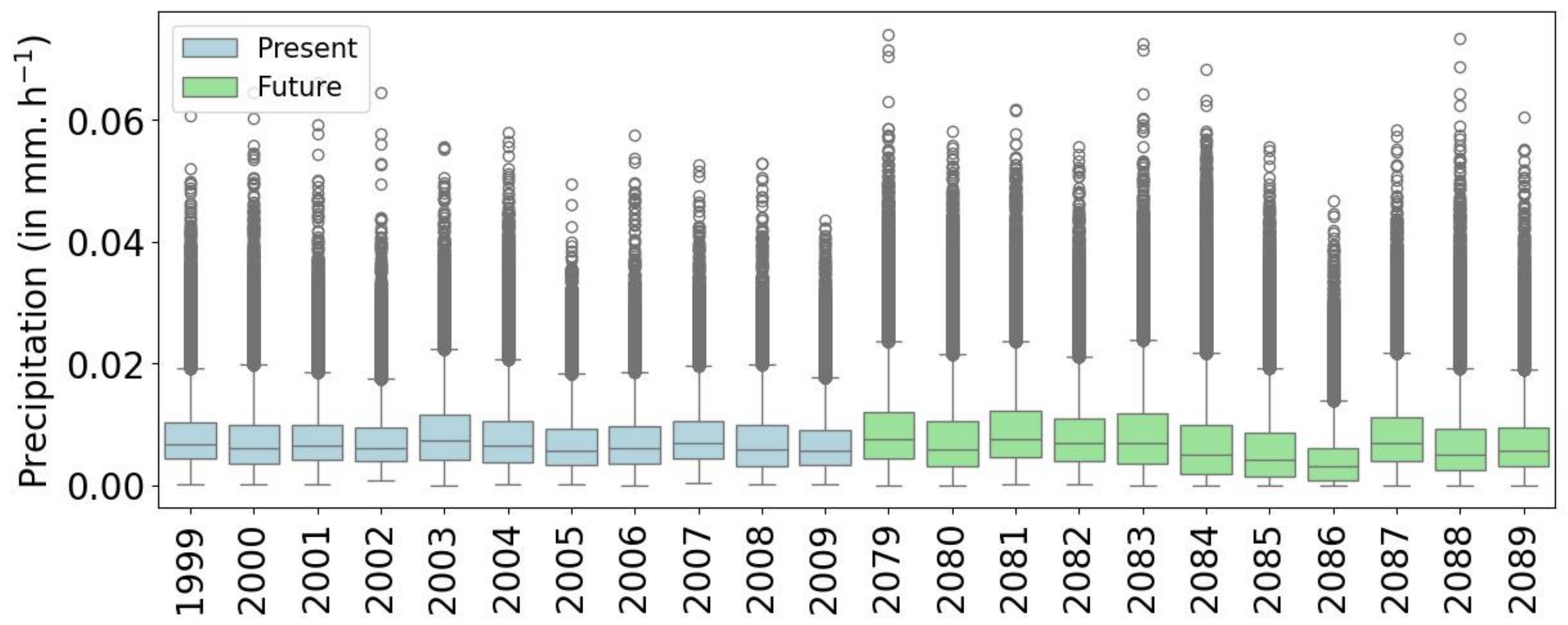}
    \caption{Distribution of monthly precipitation maxima across all grid points for each year during the 11-year present (blue) and future (green) periods.}
    \label{fig:boxplot_stationarity}
\end{figure}

Figure \ref{fig:boxplot_stationarity} shows no apparent temporal trend in either the location or the dispersion of the extremes. To ensure full rigor, we apply formal statistical tests, described in the following subsections.

\subsection{Mann--Kendall Trend Test}

Let $\{X_t\}_{t=1}^n$ denote a time series of block maxima. The Mann--Kendall (MK) test evaluates the null hypothesis of no monotonic trend against the alternative of a monotonic increase or decrease, without assuming any specific distribution for $X_t$. It is based on the statistic
\begin{equation}
S = \sum_{i<j} \operatorname{sign}(X_j - X_i),
\end{equation}
which summarizes the balance of positive and negative pairwise differences. For sufficiently large sample sizes, $S$ is approximately normally distributed with mean zero and a variance that accounts for tied values, and it is closely related to Kendall's $\tau$ coefficient.

The test is robust to outliers and does not assume linearity, making it particularly suitable for extreme precipitation series, where rare events can strongly influence trends. Its non-parametric nature ensures reliable detection of monotonic changes in block-maxima precipitation data.


\subsubsection*{Results}

\begin{figure}
    \centering
    \includegraphics[width=0.8\textwidth]{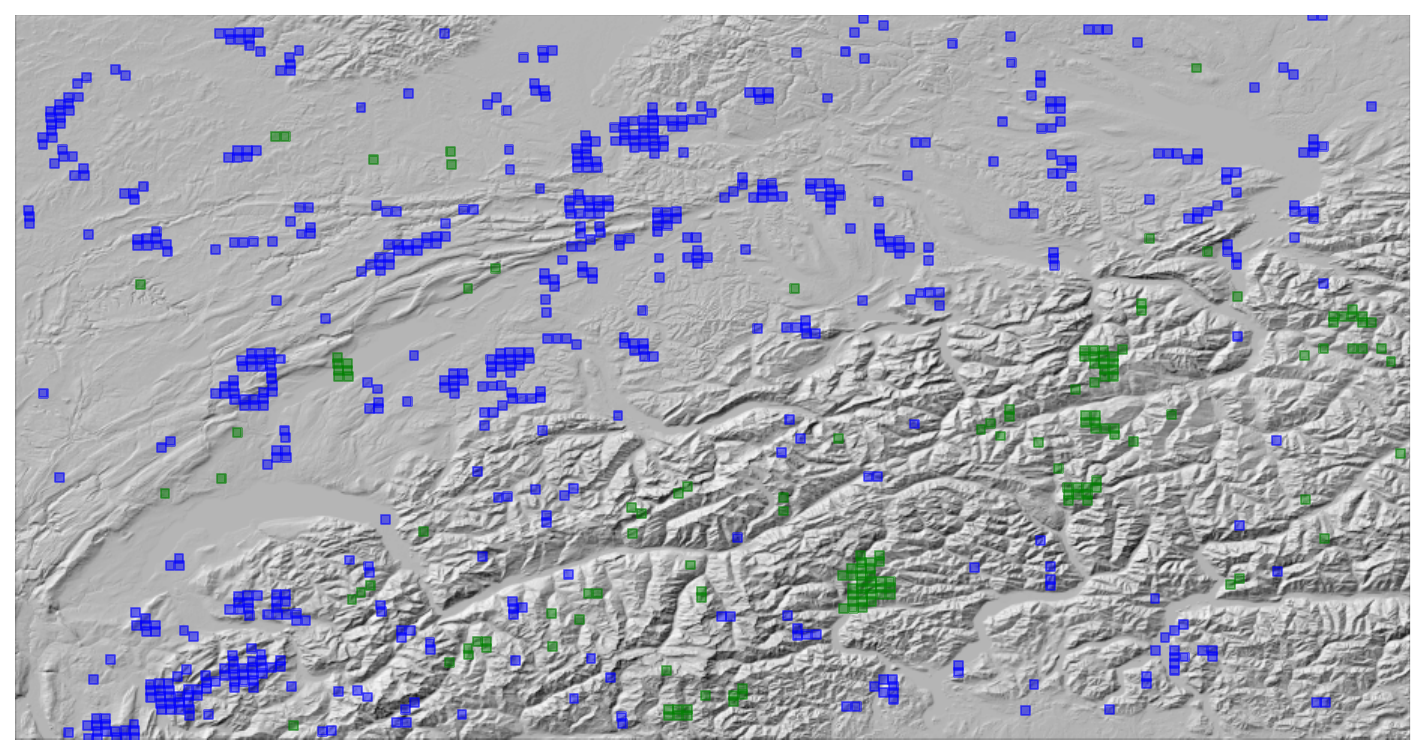}
    \caption{Maps of the pointwise Mann–Kendall test results across all grid points. Each pixel corresponds to a single grid point, with blue indicating a significant positive trend and green indicating a significant negative trend. Grid points where the null hypothesis is not rejected (no significant trend) are shown in grey.}
    \label{fig:MKtest}
\end{figure}

The MK test was applied independently at each fine-resolution grid point ($2.2$\,km grid spacing) for the 33 values of the present period. Figure~\ref{fig:MKtest} illustrates the spatial distribution of the test results, showing that only 5.25\% of grid points reject the null hypothesis at the 5\% significance level. This indicates that the vast majority of the domain showed no detectable monotonic trend in block maxima over the 11-year period.

\subsection{Likelihood-Ratio Tests for Time-Varying GEV Parameters}

To conduct a more insightful temporal stationarity analysis for our model, we test whether a significant temporal trend is present in the GEV parameters. Specifically, for each fine-resolution grid point ($2.2$\,km grid spacing), we compare a stationary GEV model with constant parameters $(\mu,\sigma,\xi)$ against models in which either the location parameter $\mu$ or the scale parameter $\sigma$ varies linearly with time.

Because each year provides three block maxima, time is encoded as
\begin{equation}
t = 1,1,1,2,2,2,\dots,11,11,11,
\end{equation}
yielding one index per year.

For each grid point and parameter, the Likelihood Ratio Test (LRT) assesses the null hypothesis $H_0$ that the parameter remains constant over time against the alternative that it varies linearly with $t$. Let $\ell_0$ and $\ell_1$ denote the maximized log-likelihoods under the stationary and time-varying models, respectively. The likelihood-ratio statistic is defined as
\begin{equation}
\Lambda = 2(\ell_1 - \ell_0),
\end{equation}
which follows a $\chi^2_1$ distribution under the null hypothesis $H_0$.

\subsubsection*{Results for the location parameter $\mu$}

The LRT showed that only a small fraction of grid points exhibited a significant time-varying location parameter. Around 95\% of the grid cells did not reject the null hypothesis, indicating that $\mu$ is effectively stationary over the 11-year period.

\begin{figure}
    \centering
    \includegraphics[width=0.8\textwidth]{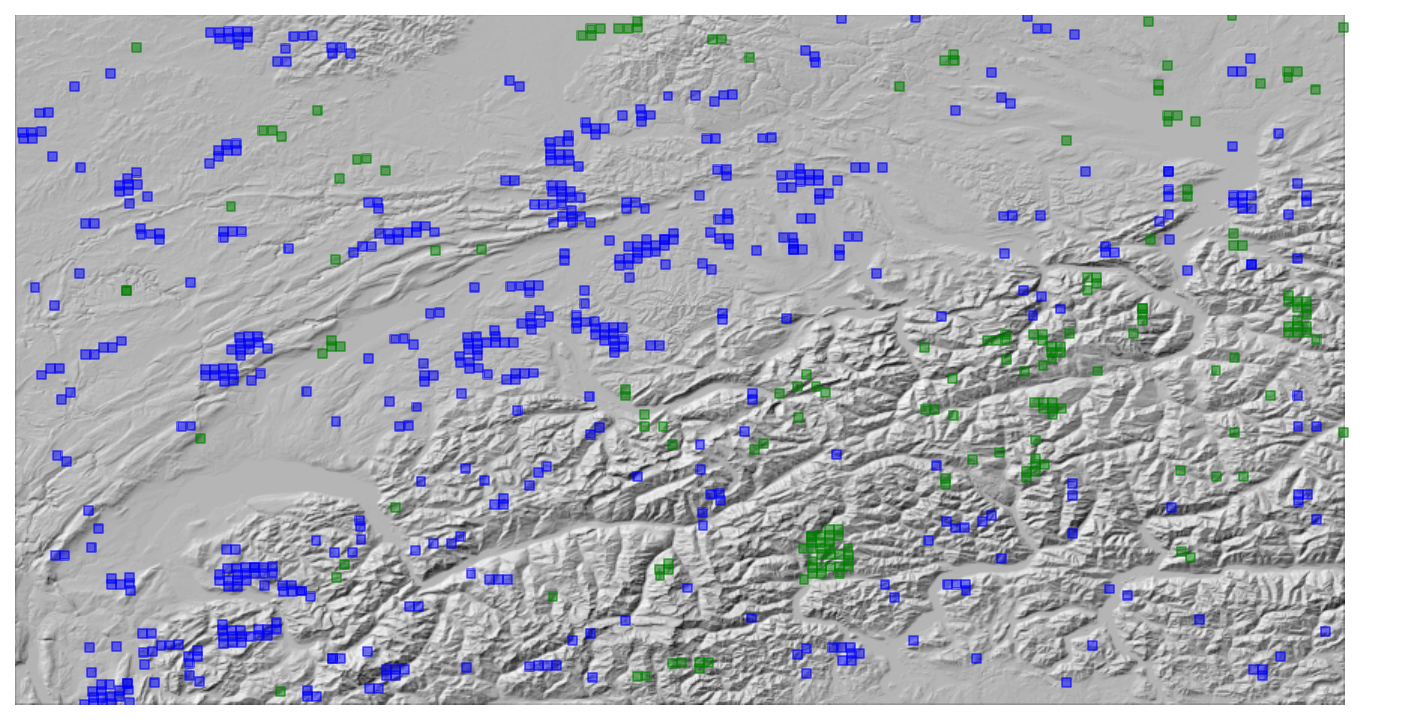}
    \caption{Maps of the pointwise LRT test results across all grid points. Each pixel corresponds to a single grid point, with blue indicating a significant positive trend and green indicating a significant negative trend. Grid points for which the null hypothesis of no trend is not rejected are shown in grey.}
    \label{fig:robustness_gap}
\end{figure}

\subsubsection*{Results for the scale parameter $\sigma$}

Similarly, the LRT applied to the scale parameter $\sigma$ (Figure~S3) shows that roughly 95\% of the grid points fail to reject the null hypothesis. Thus, the dispersion of the block maxima values does not exhibit significant temporal evolution.

\begin{figure}
    \centering
    \includegraphics[width=0.8\textwidth]{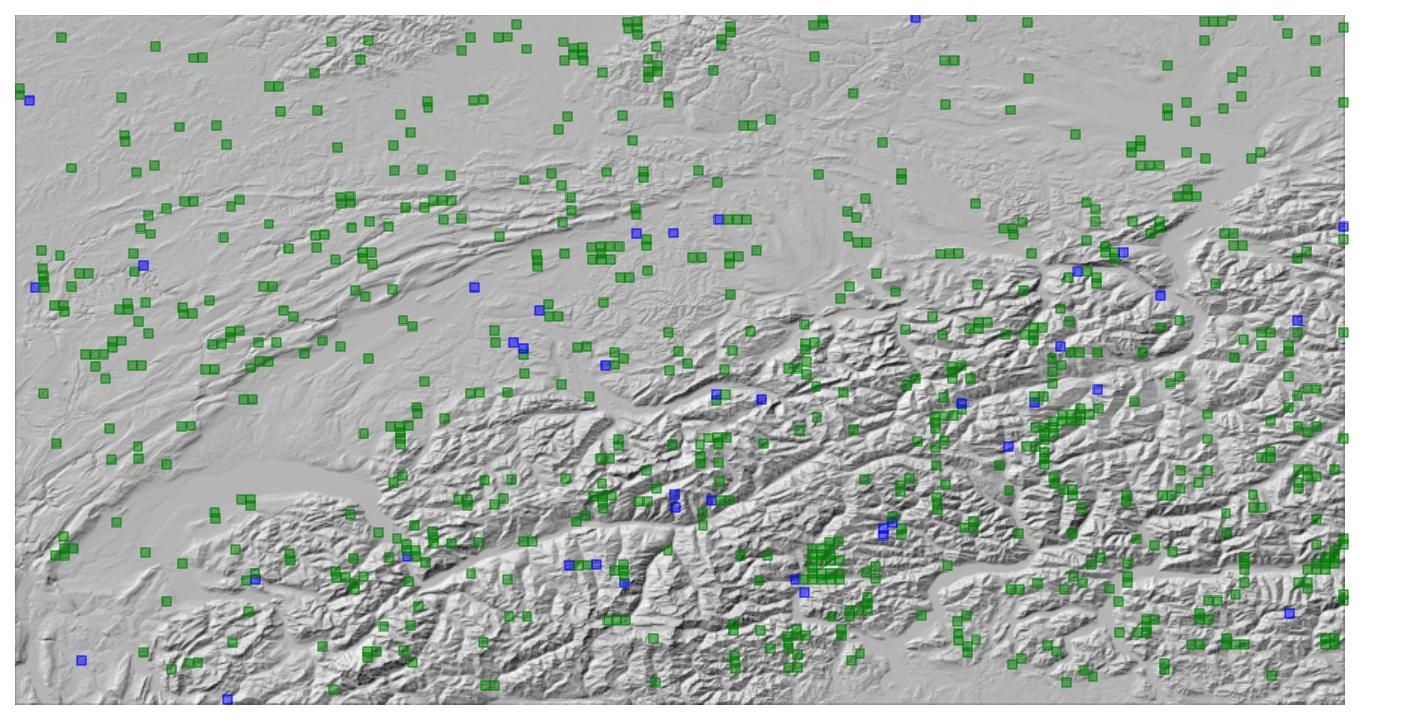}
    \caption{Maps of the pointwise LRT test results across all grid points. Each pixel corresponds to a single grid point, with blue indicating a significant positive trend and green indicating a significant negative trend. Grid points where the null hypothesis of no trend is not rejected are shown in grey.}
    \label{fig:robustness_gap}
\end{figure}

\subsection{Conclusion}

Across all three tests—the Mann--Kendall trend test and the LRTs for $\mu(t)$ and $\sigma(t)$—the vast majority of grid points showed no significant temporal dependence. The proportion of significant results would be even smaller if adjustments for multiple testing were applied. This supports the use of a stationary GEV distribution at each grid point. Similar results were obtained for the future period.

\bibliographystyle{elsarticle-harv} 
\bibliography{sample}